# 4 Toward a Generalized Theory of Observers

*Hatem Elshatlawy, Dean Rickles, and Xerxes D. Arsiwalla*

## 4.1 Introduction: Historical perspectives on observers

*The observer, when he seems to himself to be observing a stone, is really, if physics is to be believed, observing the effects of the stone upon himself. Thus science seems to be at war with itself: when it most means to be objective, it finds itself plunged into subjectivity against its will.*

–Bertrand Russell

The concept of an *observer* has been a core puzzle from ancient philosophy to modern physics, shaping how we interpret measurement, reality, and knowledge. Classical thinkers such as Aristotle placed the observer as a passive spectator to the natural world. But the beginning of modern science came with the demand for the observer to take a more active role, as expressed by Francis Bacon in his New Organon (98th Aphorism Concerning the Interpretation of Nature):

> In the business of life, the best way to discover a man's character, the secrets of how his mind works, is to see how he handles trouble. In just the same way, nature's secrets come to light better when she is artificially shaken up than when she goes her own way.

In modern physics—with the stark differences between reference frames in Einstein's relativity and with Heisenberg's uncertainty principle—the active and constraining role of the observer came to be seen as not just expedient, but as unavoidable [1, 2].

In quantum mechanics, the observer problem became explicit, ultimately raising questions about consciousness, apparatus, and whether physical law alone dictates the outcome of measurements [4, 5]. As measurement seemingly "collapses" the wavefunction, intervention was recognized as a prerequisite of observation. In the field of cybernetics, a complementary development took place: observation was clearly recognized as a prerequisite for intervention, as observation (of outcomes) provides the input to adjust the actions of an agent in a feedback loop [6, 7]. A similar development can be observed in the fields of biology [8] and artificial intelligence [9], where the observer is increasingly regarded as a feedback-driven agent or subsystem with the capacity to influence and be influenced by its environment.

This chapter aims to consolidate these diverse threads into a generalized observer theory. Specifically, we propose a *formal* definition of minimal observers, discuss their philosophical significance, and explain which concepts (such as measurement, internal vs. external distinctions, and emergent complexity) hinge on the presence of observers. We also connect these ideas to pressing questions in quantum gravity and digital physics [11, 12, 32, 33], where the ultimate structure of reality and computation may depend on observer-oriented frameworks. In addition, we point readers to dual-aspect monism arguments [14] and debates on consciousness that further highlight how observer-internal processes and external meaning may be intertwined.

Furthermore, by a *minimal observer*, we refer to the simplest possible entity that exhibits the core characteristics of observation: the ability to perceive external states, update internal configurations based on input, and generate an action or output, thereby forming a closed feedback loop. This definition, which we formalize in the following sections, ensures that the observer remains functionally distinct from its environment while engaging in meaningful interactions that shape both its perception and responses.

To make our account as rigorous as possible, we include new mathematical developments: theorems on *observer equivalence* and *observational complexity*, explicit *diagrams* to clarify feedback loops, and in-depth comparisons with established theories in physics and philosophy.

## 4.2 A general theory of observers

### *4.2.1 The need for a unified model*

Despite the variety of fields referencing "observers," there does not exist a single unifying formalism of observers. Quantum physicists define observers in terms of measurement apparatus or "external" classical systems [3, 13], while cognitive scientists see observers as perceiving agents [14], and computer scientists think of them as abstract data-collecting subroutines [10]. This
domain-based fragmentation of ideas obscures shared principles and hinders cross-disciplinary integration. A generalized observer theory seeks to: identify core *functional* features of observation (sensing, state updating, responding); clarify how these features scale from minimal feedback loops to conscious or socially embedded observers; provide a framework to address both *foundational* questions (quantum measurement, realism vs. anti-realism) and *practical* ones (AI design, ethics, interpretability).

#### *4.2.1.1 Record-keeping and information processing in observer models*

A promising avenue for unification lies in recognizing a shared invariant: *observation as an information-recording process*. Hugh Everett's formulation of quantum mechanics conceptualizes observers as *servomechanisms*, automatically functioning machines that register environmental interactions via memory storage. He proposed that observation is best understood as a record-keeping process—observers are not external agents but physical subsystems that *store measurement outcomes* as part of their own state evolution [42]. This insight suggests that observation, at its core, entails a feedback loop in which perception updates an internal record, shaping subsequent responses.

This perspective re-emerges in James Hartle's notion of *Information Gathering and Utilizing Systems* (IGUSs), a generalized framework for modeling observers in physics. IGUSs encompass any system—biological, mechanical, or computational—that collects, stores, and processes information to make decisions or generate outputs [43]. A key feature of IGUS models is their time-sequenced memory, which retains past inputs to inform future states, mirroring Everett's record-keeping servomechanisms. As Bacciagaluppi notes, Everett's concept of servomechanistic observers resurfaces in decoherence-based discussions, where IGUSs formalize the continuous accumulation and processing of data [45]. This suggests that *recorded information*—not subjective awareness—defines an observer's role in physics.

#### *4.2.1.2 Toward a unified observer framework*

By highlighting the importance of *record-keeping* across physics and cognitive models, we can extract a fundamental principle of observation. Whether in quantum measurement, artificial intelligence, or biological perception, observers can be described as information-processing units with internal states that encode and update representations of their environment. This shared functionality provides a compelling foundation for a cross-disciplinary observer theory, allowing a unified framework to emerge that reconciles disparate approaches under a common structural paradigm.

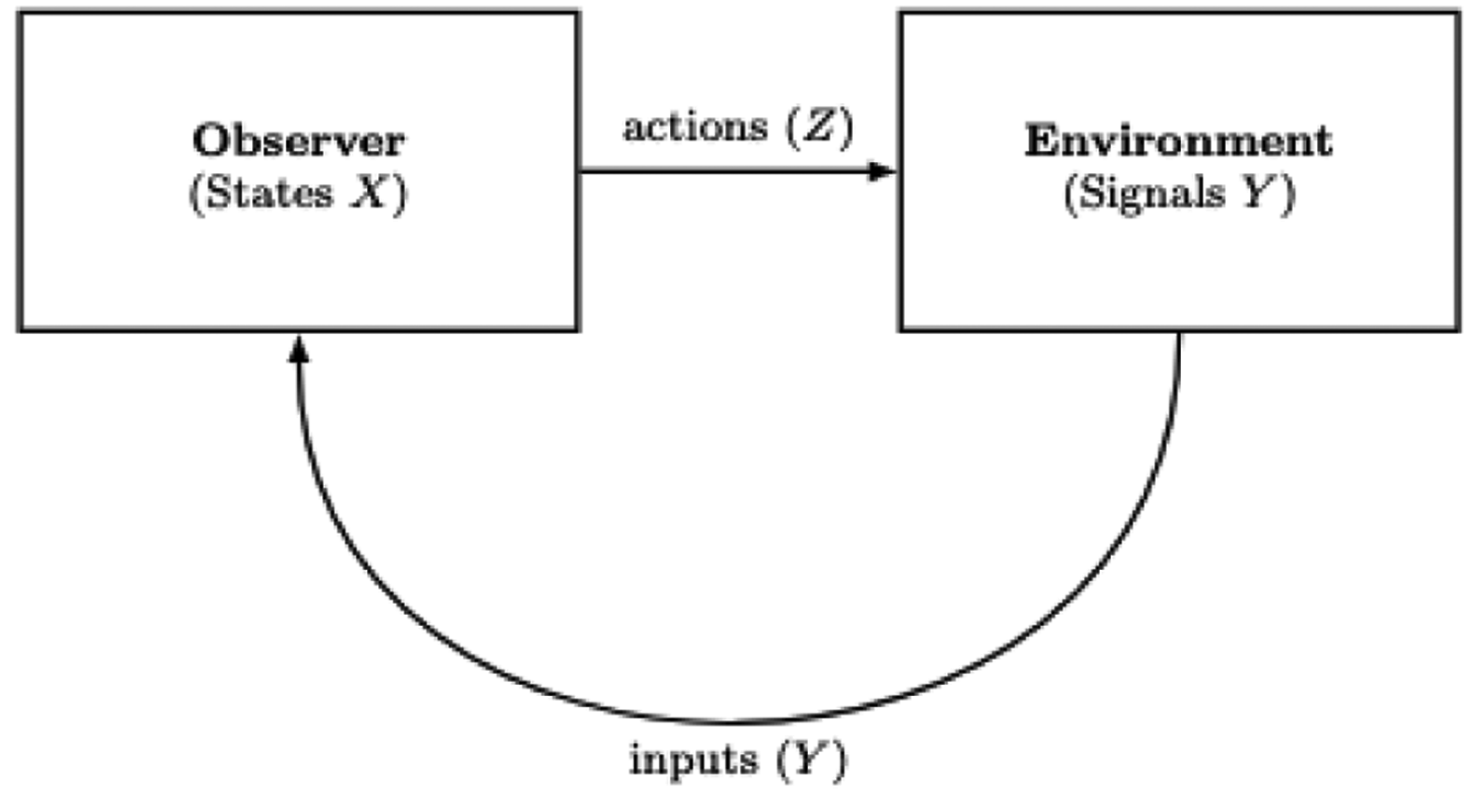

*Figure 4.1* Minimal observer-environment feedback loop.

## 4.3 Minimal observation models in cybernetics

### *4.3.1 Introduction to cybernetics and observation*

Cybernetics, as pioneered by Norbert Wiener [6] and developed by Ross Ashby [7], is the study of communication and control in organisms and machines. Its hallmark is the *feedback loop*, in which a system senses some variable, processes that information, and acts upon its environment, thus influencing future sensory input. Figure 4.1 shows a simple diagram of this loop, highlighting the observer's role.

### *4.3.2 The sensor-actuator feedback loop*

A simplified cybernetic observer consists of three essential components:

1 **Sensor:** Detects environmental or internal states (e.g., temperature, light intensity).

2 **Processing unit:** Interprets sensor data, often by comparing it to a goal or reference.

3 **Actuator:** Executes an action that changes either the system itself or its environment.

### *4.3.3 Examples of minimal cybernetic observers*

**THERMOSTAT**

A *thermostat* maintains temperature by comparing a set-point to the current temperature sensor reading and toggling a heater. Despite its simplicity, it is widely regarded as a minimal observer system: it senses (temperature), processes (comparing to a set-point), and acts (turning heating on/off).[1]

**SIMPLE REACTIVE AGENTS (BRAITENBERG VEHICLES)**

*Braitenberg vehicles* [17] demonstrate how purely reactive sensor-actuator links can yield emergent "intelligent-looking" behaviors. Although lacking complex cognition, they meet basic observer criteria by receiving sensory data, updating motor outputs, and influencing their environment in a feedback loop.

## 4.4 Second-order cybernetics

### *4.4.1 Overview*

First-order cybernetics keeps the observer *outside* the system: it is a third-person perspective. In *second-order cybernetics* (Fig. 4.2) [15], the observer is integrated *into* the system, allowing for self-reference, thus giving not just a first-person perspective, but one in which the observer enters their own domain.

### *4.4.2 Key concepts*

**Observer inclusion:** The observer is a part of the feedback process, not a neutral external vantage point.
**Self-reference:** The observer can observe and modify their own rules, leading to learning or adaptation.
**Constructivism:** Reality emerges through the observer's activities and interpretations [8, 16].

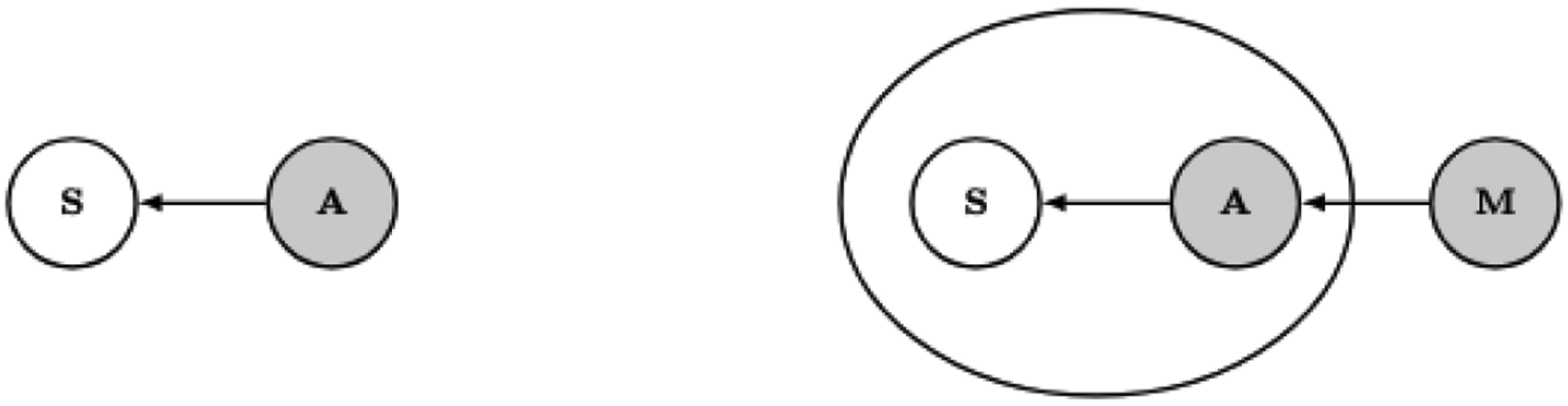

*Figure 4.2* First-order vs. second-order cybernetics: (left) first-order observer observes the system, (right) second-order observer modifies the observation process itself.

**EXAMPLE: LEARNING NEURAL NETWORKS**

A neural network updating its weights $W$ in response to error signals exemplifies second-order cybernetics. The network *observes* (via a loss function) the mismatch between outputs and targets, adjusting $W$ to reduce future mismatch—thus reconfiguring its own internal parameters.

## 4.5 Developing a meta-model for observers

### *4.5.1 Core components of the meta-model*

Across many examples, we identify the following recurring features:

1 Sensing mechanism (perception)

2 Processing unit (interpretation)

3 Response mechanism (action)

4 Feedback loop (adaptation)

5 Internal model or representation (prediction)

6 Boundary definition (self vs. environment)

7 Self-monitoring (self-observation)

### *4.5.2 A formal definition of minimal observers*

**Definition 4.1 (minimal observer):** Let $O$ be a system described by the tuple

$$O = (X, Y, Z, f, g, \mathcal{B}),$$

where $X$ is internal state space (finite or countably infinite), $Y$ is input (sensor) space, $Z$ is Output (action) space, $f$ is $X \times Y \to X$: State transition function, $g$ is $X \to Z$: Output function, and $\mathcal{B}$ is a boundary condition demarcating "inside" (the observer's internal states) vs. "outside" (the environment).

Then $O$ is *minimal* if:

$|Y| \geq 1$ (non-trivial sensing),

$|Z| \geq 1$ (non-trivial action),

$|X| > 1$ (non-trivial internal dynamics),

Feedback closure: the observer's actions $g(x)$ alter the environment, which in turn alters subsequent inputs $y \in Y$.

#### WHAT DOES "MINIMALITY" MEAN HERE?

This definition positions the minimal observer as the simplest system capable of observation in a functional sense: it must sense ($Y$), update its state ($f$), and act ($g$), with a boundary ($\mathcal{B}$) and feedback loop ensuring interaction with an environment. This is minimal in a structural sense—it avoids complexity like memory, learning, or self-awareness, focusing on the bare essentials of observation. However, "minimality" could be interpreted in multiple ways: **structural minimality**, referring to the fewest components needed for observation (e.g., the thermostat example below (Section 4.5.3) fits structural minimality: it exemplifies a basic feedback system with no self-modification or self-production, both of which lie beyond the scope of this minimal model); **functional minimality**, denoting the simplest system that still performs a meaningful role (e.g., distinguishing internal vs. external states); and **ontological minimality**, which identifies the foundational unit from which all observer-like phenomena emerge (possibly tying to autonomy or autopoiesis).

### *4.5.3 Thermostat as a formal example*

As an illustration, let $X = \{\text{ON}, \text{OFF}\}$, $Y = \{\text{Cold}, \text{Hot}\}$, $Z = \{\text{HeaterOn}, \text{HeaterOff}\}$. Define

$$f(\text{OFF}, \text{Cold}) = \text{ON}, \qquad f(\text{OFF}, \text{Hot}) = \text{OFF},$$
$$f(\text{ON}, \text{Cold}) = \text{ON}, \qquad f(\text{ON}, \text{Hot}) = \text{OFF}.$$

and

$$g(\text{ON}) = \text{HeaterOn}, \quad g(\text{OFF}) = \text{HeaterOff}.$$

A boundary $\mathcal{B}$ physically partitions the controller from ambient air. This meets minimal observer criteria: the thermostat senses "Cold/Hot," toggles $\{\text{ON}, \text{OFF}\}$, and acts by turning heat on/off.

## 4.6 Foundational questions and observer-dependent concepts

### *4.6.1 Relating the model to foundational questions*

#### MEASUREMENT IN PHYSICS AND OBSERVER ROLES

Quantum theory famously hinges on measurement, prompting debates over whether wavefunction collapse is triggered by consciousness, classical apparatus, or decoherence [4, 5, 13]. Our minimal observer model, although classical, illuminates the *functional* demands of measurement: a system must sense, store, and act upon the input, forging a boundary that designates what is measured (the environment) and what is measuring (the observer). This boundary-centric perspective resonates with relational interpretations of quantum mechanics, where all states and events are observer-relative [28].[2]

#### COMPUTATION AND COMPLEXITY

Turing machines [10] epitomize minimal universal computation. Our minimal observer is "less ambitious": it does not demand universal problem-solving but ensures a fundamental sensor-actuator feedback. Nevertheless, bridging these models can highlight interesting points, such as whether an observer can, in principle, simulate arbitrary computational processes if given enough states $X$ and a suitable set of transitions $f$.

#### CONSCIOUSNESS DEBATES

Whether minimal observers can illuminate the "hard problem" of consciousness [7, 12, 18, 19] remains debatable. Yet the boundary $\mathcal{B}$ and internal modeling $h$ in more complex observers may underlie self-referential processes that are often considered key to subjective experience and conscious phenomenology [14, 20–27]. While minimal observers do not *entail* consciousness, they provide building blocks to analyze how layered observation might scale up to phenomena associated with awareness.

### *4.6.2 Identifying what would not be defined without observers*

Without an observer:

1. There is no clear distinction between **internal vs. external space.**
2. There are no definitive **measurement outcomes.**
3. There are no **reference frames or contextual frameworks.**
4. There is no layering of **hierarchical observation** (e.g., organizations, societies, or multi-level apparatuses).
5. There is no **coarse-graining or hierarchical organization**, since it is observers who impose boundaries that allow phenomena to be described at different scales or levels of abstraction.

*Observers provide the partitions that make physics (and meaning) possible.* This point resonates strongly with the "ruliological" viewpoint of Stephen Wolfram: in his computational universe picture, reality only becomes tractable once an observer chooses a particular foliation of the underlying causal network, thus creating the very coarse-grained structures, frames, outcomes and hierarchical layers, that we then treat as the fabric of experience and scientific explanation [60, 66–68] (Fig. 4.3).

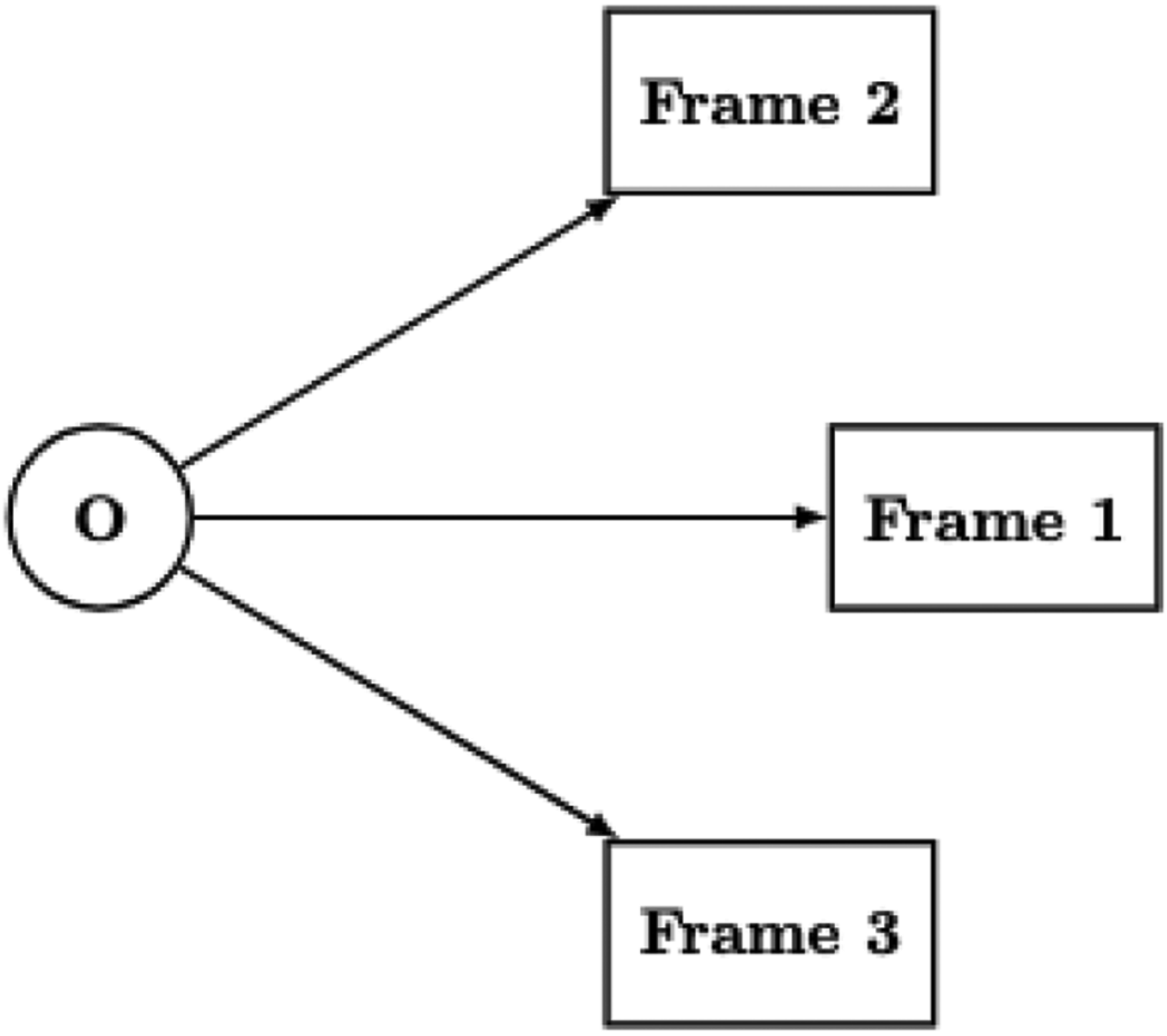

*Figure 4.3* Observer defining multiple reference frames in physical or cognitive space.

Hierarchical structures, as mentioned earlier, rely fundamentally on observers who define internal states and boundaries at multiple scales. Observers thus serve as "interfaces" that structure interactions between systems and their environments. This perspective resonates strongly with Chris Fields' work, which highlights how boundaries and interfaces created by observers define informational interactions between system components and their surroundings [46]. Without observers, there would be no principled way to partition reality, establish hierarchical frameworks, or define coarse-grained descriptions that enable meaningful layers of interpretation to emerge.

Hence, observation acts not just as a mechanism but as a conceptual foundation for partitioning, measuring, and categorizing phenomena.

## 4.7 Implications and insights

### *4.7.1 Case study: Is an electron an observer?*

One recurring question is whether fundamental particles (e.g., electrons) qualify as observers. By Definition 4.1, an electron does not meet the minimal criteria. While it interacts with fields, there is no *internal mechanism* that updates "electron states" based on measured input. Instead, quantum mechanics describes its evolution via the Schrödinger equation or quantum field interactions, not a sensor-actuator model with an internal feedback loop. The electron is *observed* but does not itself *observe*, lacking a definable boundary $\mathcal{B}$ that separates "internal states" from "environmental data" in a cybernetic sense.

### *4.7.2 Quantum gravity and digital physics*

In approaches to quantum gravity and pregeometric physics [11, 30, 32–34], space time emerges from underlying discrete structures or informational processes [35–41]. If the universe is essentially computational [12, 31], then observers play a crucial role in defining events or discrete state updates. Specifically, minimal observers could serve as *anchors* that stabilize local measurements, effectively converting "potential states" into classical-like outcomes. They might also define local reference frames or regions of emergent geometry, introducing boundaries into an otherwise unbounded computational cosmos and thus giving localized meaning to information flows.

These considerations align closely with constructivist perspectives linking formal languages and information structures to emergent physical realities [33]. Such minimal observer models not only enrich debates on how classicality or geometry arises from more fundamental substrates but also have implications for philosophical frameworks like dual-aspect monism [14]. Notably, the minimal observer concept might provide fresh insights into long-standing debates concerning the relationship between observers, consciousness, and the fundamental structure of reality.

Indeed, the role of observers as definers of boundaries and informational interfaces offers a bridge to epistemic or informational interpretations of quantum mechanics (such as QBism) and aligns with dual-aspect monism, where mental and physical properties emerge simultaneously from underlying informational substrates [14]. Thus, the notion of minimal observers might serve as a unifying foundation, providing a concrete model to explore how the dual aspects of subjective experience and objective reality might co-arise from a more primitive, informational substrate. As noted by Bacciagaluppi [45], this view can be traced back to Everett's treatment of observers, further suggesting a coherent synthesis of quantum epistemology, constructivism, and digital physics.

## 4.8 Computational models of minimal observation

### 4.8.1 Cellular automata and universality

#### RULE 110 AND BEYOND

Rule 110 is a one-dimensional cellular automaton that, despite having extremely simple local rules, is known to be Turing-complete [12]. However, to see it as an *observer*, we must specify how certain cells (or patterns) "sense" local configurations and produce "outputs" that affect the environment. Within such automata, a minimal observer can be implemented as a sub-lattice that monitors local states and changes them according to a rule $f(x, y)$. This demonstrates how emergent complexity might arise from repeated, local observation-based updates (Figs. 4.4–4.6).

### 4.8.2 Synthesizing with the meta-model

When inserted into a larger computational or physical context, the minimal observer definition acts like a module. For instance, in a swarm of robots or distributed computing networks, each node or agent can be viewed as a minimal observer with states, sensor channels, and outputs. By chaining or nesting these observers, we can analyze how *coherent group behaviors*, consensus, or emergent patterns appear in multi-agent systems.

## 4.9 Ontology of observers in physics

### 4.9.1 Internal vs. external spaces

The distinction between internal and external spaces, and consequently the concept of a boundary, is fundamental in fields ranging from thermodynamics to field theory. Observers explicitly define the boundary $B$ that transforms unstructured "outside" data into structured, measurable signals and internal states [47]. Figure 4.7 provides a schematic of how an observer in a continuum setting might define a region of interest for measurement, resonating strongly with relational or observer-relative approaches in physics [28].

As mentioned, this boundary construction closely parallels Chris Fields' analysis of observers as information-theoretic interfaces that partition physical systems into interacting but distinct subsystems, thereby establishing the conditions necessary for objective measurement and communication [46]. Fields' interpretation highlights that the observer's choice of interface—analogous to the Heisenberg cut in quantum mechanics—explicitly delineates what parts of the universe are considered "observed systems" versus "observing apparatus," reinforcing that these distinctions are inherently observer-defined and context-dependent [46]. Thus, the act of boundary definition emerges as a central, indispensable feature of any meaningful observer model, linking classical cybernetic observers with foundational quantum-theoretic discussions.

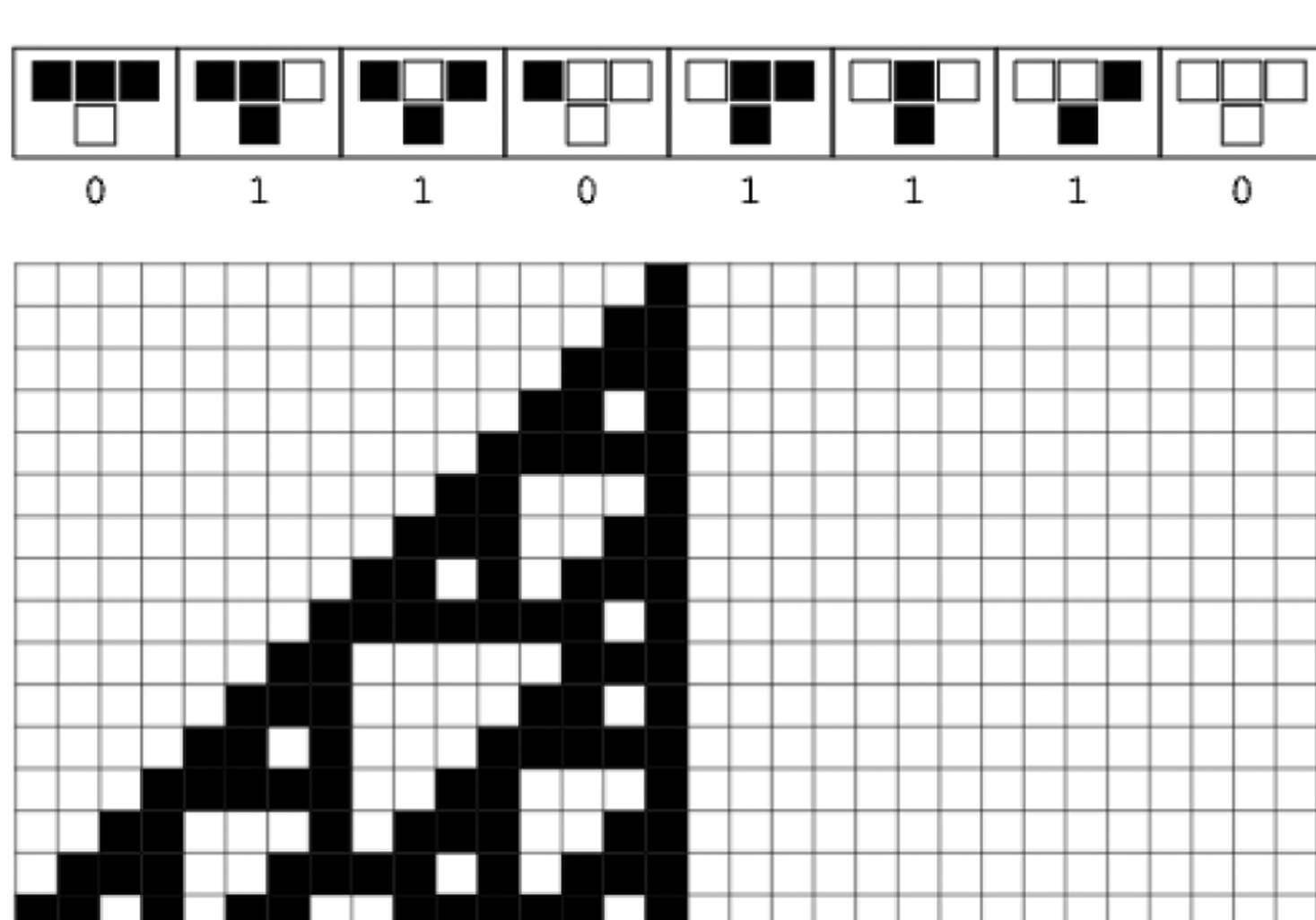

*Figure 4.4* Rule 110 is one of the *elementary cellular automaton* rules introduced by Stephen Wolfram [12]. It specifies the next color in a cell, depending on its color and immediate neighbors. Its rule outcomes are encoded in the *binary* representation $110 = 01\ 101\ 110_2$. This rule is illustrated above together with the evolution of a single black cell it produces after 15 steps.

### *4.9.2 Hierarchies of observers*

**NESTED OBSERVATION IN SOCIAL SYSTEMS**

A corporation may act as an observer by collecting data (markets, consumer feedback), processing internal states (policy decisions), and acting outward (product releases). Individual employees also act as sub-observers, forming a nested or hierarchical structure [29].

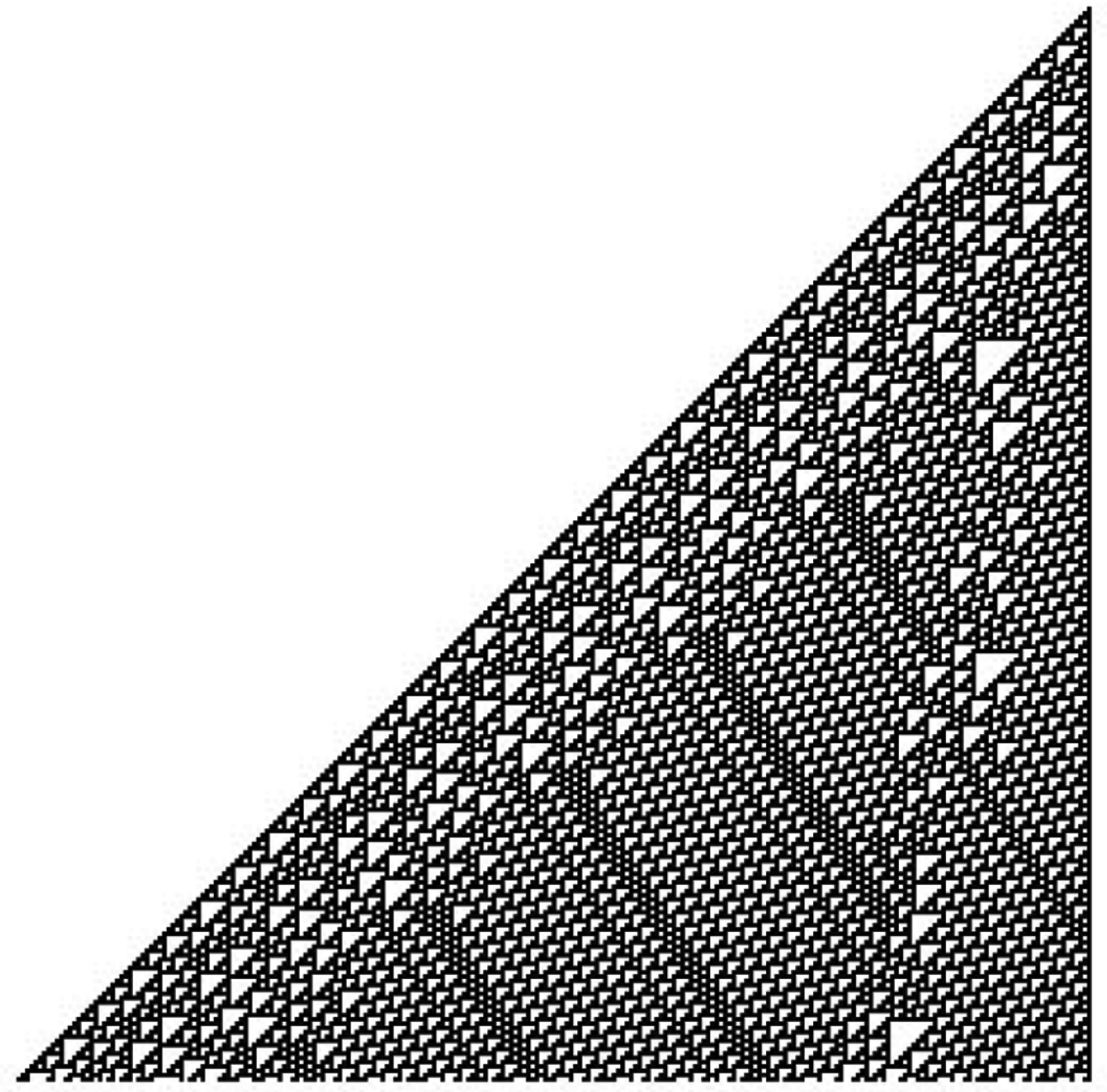

*Figure 4.5* 250 iterations of CA rule 110.

**LAYERED APPARATUS IN PHYSICS**

Large experiments (e.g., particle colliders) have multiple layers of detectors, each "observing" sub-events. Their outputs feed into aggregating devices that *observe the observers*, creating a second-order loop that yields final "measurement outcomes." An illustration of a nested observation is shown in Fig. 4.8.

## 4.10 Expanding the framework: Philosophy, theorems, ethics, and comparisons

### *4.10.1 Philosophical depth and engagement*

#### *4.10.1.1 Kant, Husserl, and Wittgenstein Revisited*

**KANTIAN TRANSCENDENTAL IDEALISM**

Kant posited that the mind actively organizes experience via innate structures (space, time, categories). Observers in our model impose a boundary $B$ and define functions $f, g$ to interpret signals, analogous to Kant's forms of intuition and categories. The environment-in-itself remains inaccessible; what the observer registers is a structured "phenomenal" realm. Formally, one might interpret $f : X \times Y \rightarrow X$ as the "transcendental function" shaping raw input $Y$ into the observer's internal "categories" $X$. This perspective deepens the epistemological stance that "raw data" cannot be known independently of the observer's interpretive structures. Ernst von Glasersfeld's radical constructivism pushes this notion further, proposing that knowledge is not representational of an external reality; rather, it emerges entirely from the observer's self-constructed experiential interface, reinforcing the idea that observers actively construct their own experiential worlds [56]. It also suggests potential alignments with neo-Kantian approaches and contemporary structural realism, where relational structures and observational interactions become more fundamental than intrinsic properties or objects themselves [57, 58].

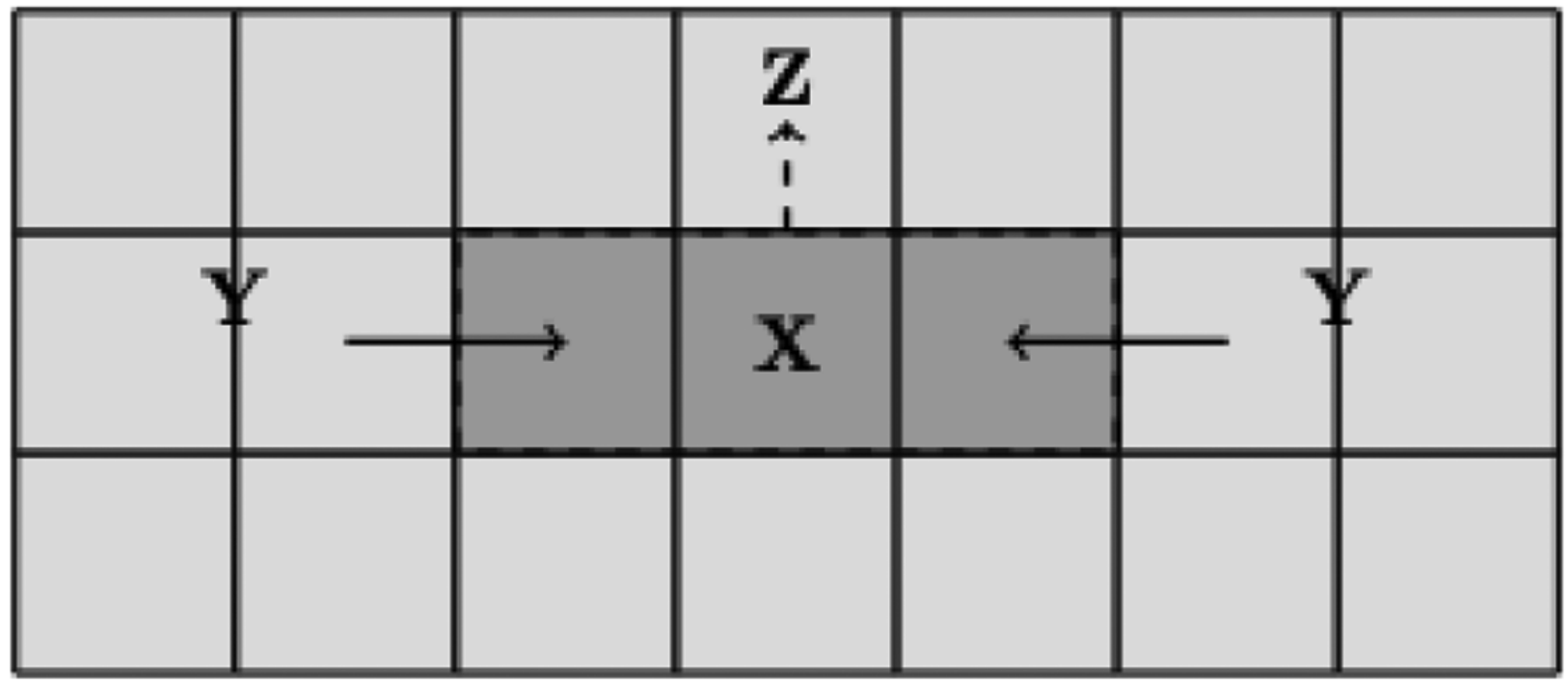

*Figure 4.6* A cellular automaton segment where a "block" of cells is designated the observer. The observer's boundary $\mathcal{B}$ encloses internal states $X$ and defines which neighboring cells constitute the input $Y$. Actions on the local environment serve as $Z$.

#### PHENOMENOLOGY AND LANGUAGE

Husserl's phenomenology and Wittgenstein's later philosophy converge on a single theme that fits naturally into our observer formalism: *meaning arises only through an observer's situated activity*. For Husserl, consciousness is *intentional*—always directed toward some object. A minimal observer's feedback loop realizes a proto-intentionality: sensor inputs $Y$ select a target state, the internal update $f$ incorporates that target, and the action $g$ projects the observer back toward its object. Husserl's *epoché* then appears as a higher-order modulation of the boundary $\mathcal{B}$, bracketing certain inputs while foregrounding others and thereby re-drawing the line between "internal'' and "external.'' Wittgenstein adds the public dimension: what counts as a meaningful output $g(x)$ depends on the "language game'' played with neighboring observers. Each observer carries a local rule-book-its particular $f, g$ pair— and different rule-books yield different semantic fields. Agreements or conflicts in multi-observer settings thus become questions about whether two feedback loops can be made partially isomorphic. In short, intentional directedness (Husserl) and rule-following meaning (Wittgenstein) are two sides of the same cybernetic coin: both reduce to how an observer carves the world with $\mathcal{B}$ and updates themselves through $f, g$ in dialogue with other agents [44].

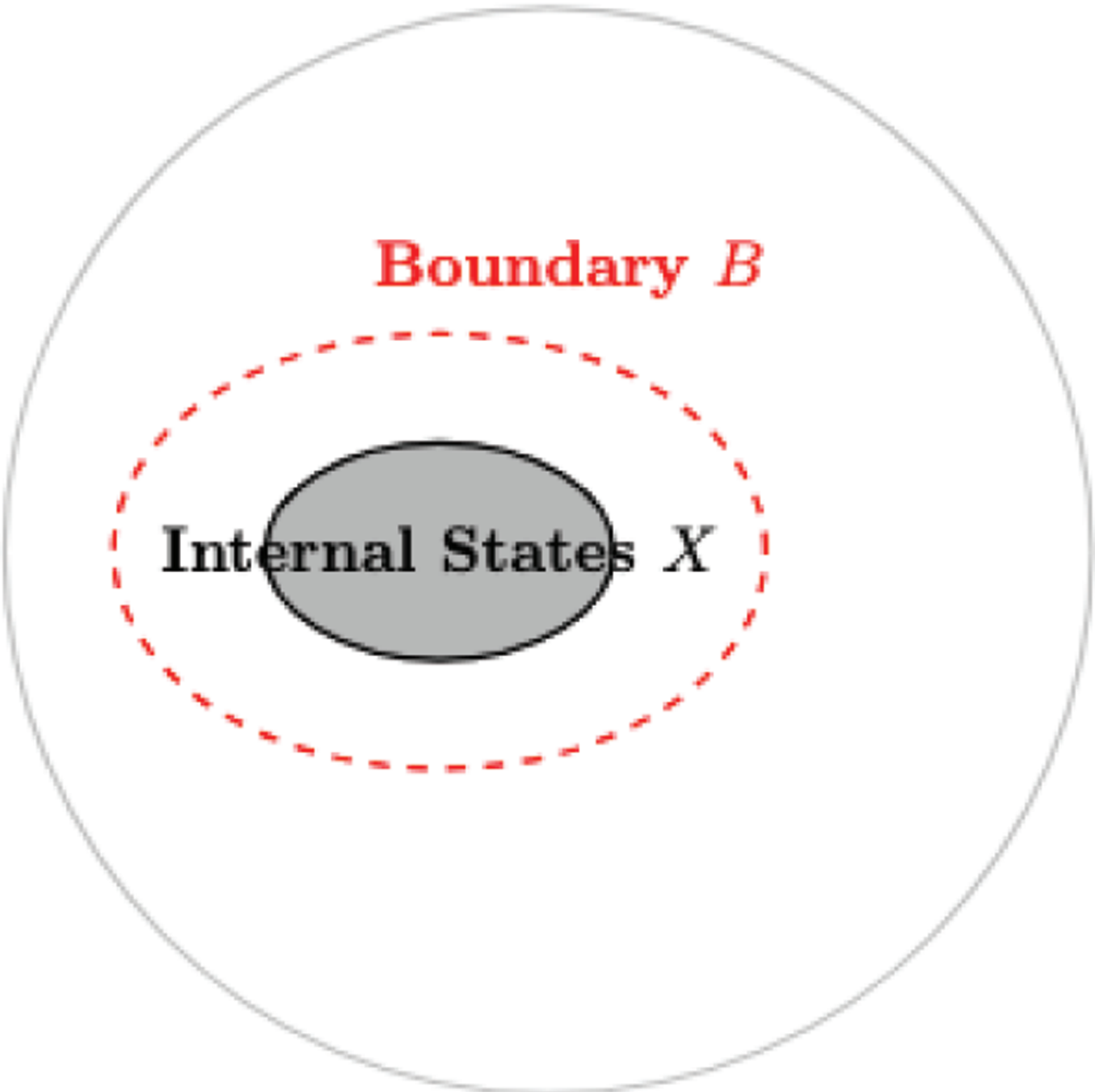

*Figure 4.7* An observer in a continuum field scenario defines a boundary $\mathcal{B}$ around certain degrees of freedom (light grey region). The external environment (white) becomes input $Y$, while internal states (dark grey) belong to $X$.

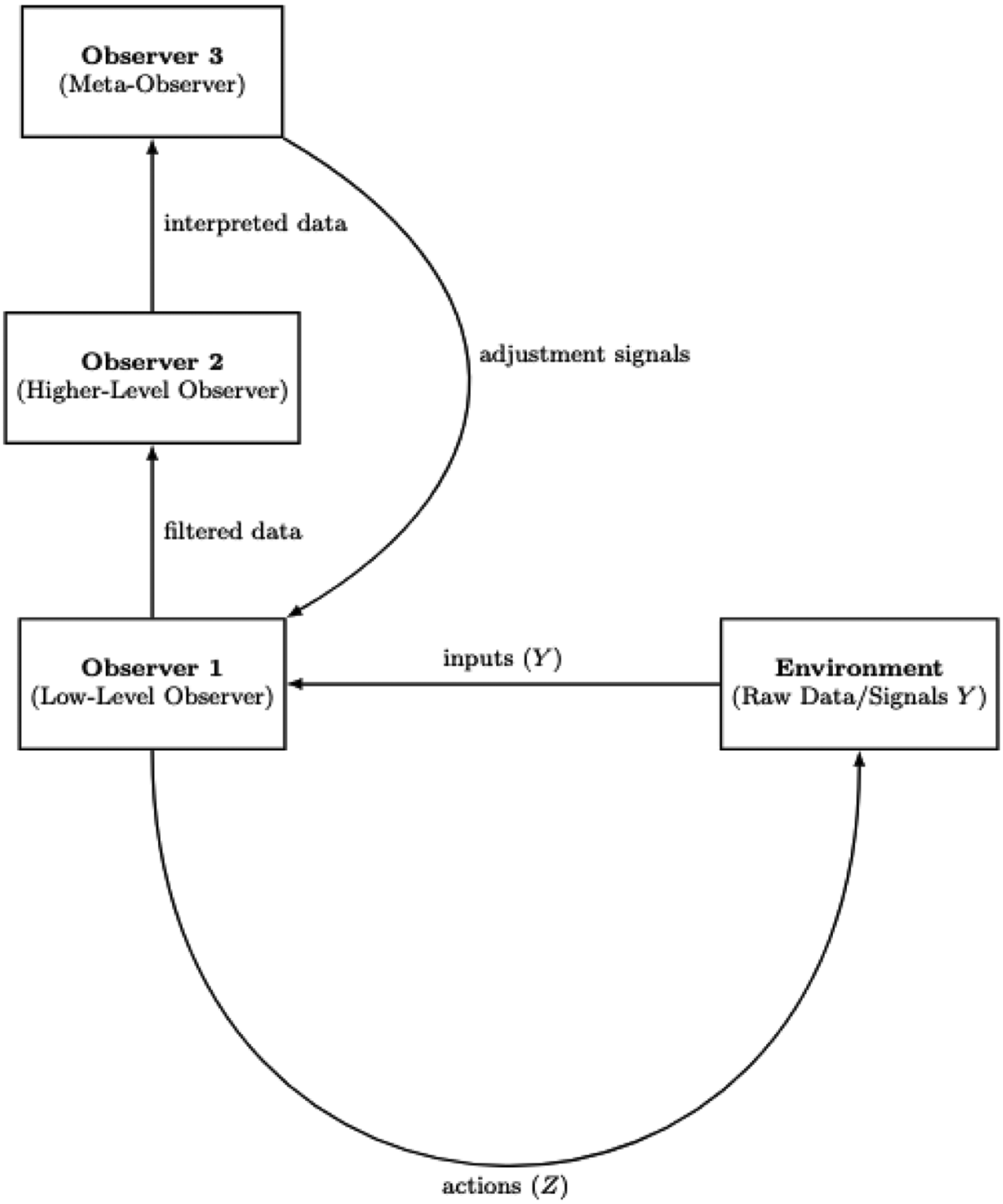

*Figure 4.8* Hierarchical observer model: higher-level observers refine and modify the interpretations of lower-level observers, creating a multi-layered observational structure.

#### *4.10.1.2 Realism, anti-realism, and constructivism*

Our formalism supports a "middle ground" between strict realism (the environment fully pre-exists observer involvement) and radical anti-realism (the observer creates all facts). Observers co-define phenomena via their action-perception loops; the environment may exist independently, but it is only "measured" or *objectified* once an observer's boundary $\mathcal{B}$ engages with it. Here, we use "measured" in a deliberately qualified sense to emphasize that observation does not merely uncover pre-existing properties but actively participates in structuring the phenomenon observed.

Bhaskar's critical realism [48] posits a stratified reality existing independently, yet acknowledges that it is only partially accessible through observation. Within our framework, the observer's internal state space $X$ explicitly quantifies how much of external reality is filtered by the observer's structure. Thus, our model enriches Bhaskar's viewpoint by providing a rigorous, formal mechanism—namely, the interplay between boundary conditions ($\mathcal{B}$) and internal states ($X$)—for understanding how observers partially shape what is perceived as reality.

Similarly, the constructivist tradition, exemplified by Maturana and Varela's autopoietic observers, views reality as actively brought forth through ongoing sensor-actuator engagement [8, 16]. Our framework adds depth to this stance by clarifying how exactly observer-boundary interactions evolve dynamically, capturing how reality emerges through a continuous historical process of state updates. It thus provides a computational underpinning to constructivist arguments, showing explicitly how perception-action loops and boundary dynamics lead to the co-creation of observer-environment distinctions over time.

In essence, our observer model bridges realism and constructivism by demonstrating concretely how structured interaction via boundaries and feedback loops can yield an objective yet observer-relative ontology, adding a formal, computational dimension to longstanding philosophical insights.

#### *4.10.1.3 Ethical implications*

##### AGENCY IN AI AND RESPONSIBILITY

When AI systems self-modify their boundaries, the question arises: *who is responsible* if such an observer redefines what data it collects or how it acts? Minimal observers who reconfigure their own $f, g, \mathcal{B}$ approach an autonomy that complicates standard accountability models. As AI systems evolve, they might not only change internal parameters (*weights*) but also alter how inputs and outputs are partitioned or categorized. This redefinition of $\mathcal{B}$—the boundary delineating "system" from "environment"—blurs traditional lines of accountability.

Questions surrounding *moral patiency* and *moral agency* become central [49–51], prompting policymakers and ethicists to reconsider frameworks for attributing responsibility or liability.

#### ENVIRONMENTAL DEFINITIONS OF VALUE

An ecological observer, as an embodied and situated agent, might sense various biodiversity metrics, interpret them via $f$, and act to conserve or exploit those measures [53–55]. The boundary $\mathcal{B}$ it adopts—be it an economic perspective or an ecological one—determines how "value" is recognized or ignored. Observers thus shape environmental policy by selecting relevant signals and discarding others. This perspective suggests a dynamic interplay between how we define "resources" or "assets" and the ways we measure them. If an observer is geared toward short-term economic gain, its $f$ function might ignore longer-term ecological consequences [52]. Conversely, a sustainability-focused observer would incorporate broader temporal scales and systemic interdependencies, thereby reshaping policy outcomes. Integrating observer theory into environmental ethics could guide more holistic decision-making processes, revealing blind spots in purely market-driven valuations.

### *4.10.2 Comparison with existing observer theories*

#### *4.10.2.1 QBism: Formalizing belief updates with an observer's boundary*

#### THE QBISM'S VIEW

Quantum Bayesianism (QBism) treats the wavefunction as an observer's subjective belief about a quantum system. Rather than describing objective system properties, quantum states reflect an agent's personal degrees of belief. Importantly, QBism does not prescribe belief updates via the Born rule as probability updating (diachronic), but rather employs the Born rule to detect inconsistencies within an agent's probability assignments at a given moment (synchronic coherence) [68]. QBism similarly does not mandate Bayesian updating as the sole coherent strategy—belief updates depend fundamentally on each agent's particular belief conditions [68, 69].

While QBism clarifies measurement interactions conceptually, it deliberately avoids providing a third-person or physical dynamical mechanism explaining how belief states physically interact with measurement outcomes. This is intentional: QBism seeks to dissolve the measurement problem by reframing quantum mechanics as a decision-theoretic tool for agents, thus avoiding explanations via another physical process [70]. Nevertheless, clarifying the agent–environment boundary remains conceptually beneficial, particularly in explicit modeling scenarios.

#### STRUCTURED OBSERVER–SYSTEM INTERACTION

Our minimal observer framework explicitly formalizes these concepts by defining an observer as a structured entity $O = (X, Y, Z, f, g, \mathcal{B})$, with internal states $X$ representing subjective beliefs, and boundary $\mathcal{B}$ explicitly distinguishing the observer from its environment. Measurement formally corresponds to sensory inputs $y \in Y$ crossing the boundary, updating the observer's internal state via a general transition function $f : X \times Y \rightarrow X$. Crucially, f need not represent solely Bayesian updating but can reflect any coherent belief-adjustment strategy consistent with the agent's prior conditions. Thus, rather than resolving ambiguities in QBism, this formalism concretizes QBist principles in an operational manner, explicitly modeling the interface between subjective belief and external quantum events.

#### PHENOMENOLOGICAL SCENARIO (QBism)

Consider a quantum coin-flip scenario. Initially, an observer encodes subjective uncertainty regarding outcomes. Upon measurement, the observed outcome crosses the boundary $\mathcal{B}$ into sensory input space $Y$, prompting state-transition $f$ to yield a new internal belief state. This concrete depiction operationalizes QBism's subjective belief perspective, explicitly modeling informational flow across agent-system boundaries. Such explicit modeling, while not required by QBism, offers clarity and enables exploration of scenarios involving multiple interacting observers.

#### PHENOMENOLOGICAL ILLUSTRATION (QBism)

Imagine two observers with distinct priors observing the same quantum coin toss. Each observer, represented by internal state sets $X$, independently updates beliefs after measurements. Subsequent communication (interactions crossing boundaries $\mathcal{B}$) allows observers to exchange information; however, alignment of subjective beliefs is not guaranteed by mere communication alone, as Bayesian agents require substantial prior agreement (e.g., overlapping priors or common sample spaces) to reach consensus [69]. This explicitly modeled observer framework thus allows rigorous and nuanced exploration of QBism's relational epistemology, emphasizing the complexity and depth inherent in subjective belief alignment.

### *4.10.2.2 Relational quantum mechanics: Observer-relative states via internal structure*

#### THE RQM VIEW

Relational Quantum Mechanics (RQM) asserts that the state of a system is not absolute; it only exists relative to a given observer or reference frame [28]. Different observers can have different accounts of a sequence of events, and there is no "God's-eye-view" wavefunction for the whole universe. In RQM, any physical interaction can play the role of an observation—even an inanimate object can be an "observer" in the sense that the object has a state relative to another system. However, RQM as originally formulated (e.g., by Carlo Rovelli) provides a conceptual framework rather than a detailed operational model: it says each observer might have their own Hilbert space of information, but it doesn't specify how an observer is structured or how exactly one defines when a fact becomes relative to a particular observer. For instance, RQM contends that if observer $A$ measures system $S$, then $S$ has a definite outcome state for $A$, but a second observer $B$ who hasn't interacted may still describe the composite $A + S$ as in a superposition. This raises the question: what precisely counts as an "observer," and what is the criterion for an event being realized relative to that observer? The minimal observer model addresses this by giving a rigorous operational definition of an observer and the moment when something becomes a fact for that observer.

#### FORMALIZING THE RELATIONAL ROLE OF OBSERVERS

In our model, the boundary $\mathcal{B}$ explicitly delineates the division between an observer's internal degrees of freedom and the external system. This provides a concrete way to implement RQM's core idea that "state is relative to the observer." According to the model, a physical interaction becomes an observation when some input $y$ crosses $\mathcal{B}$ and updates the internal state $X$ of the observer. At that moment, we say the external system has a definite property relative to that observer, encoded by the observer's internal record. Each observer $O = (X, Y, Z, f, g, \mathcal{B})$ thus sees a different state of the world, as each maintains its unique internal record reflecting its interaction history. Our framework makes this relational nature explicit: if a quantity about the system is not encoded in $X$ (because the interaction hasn't occurred), then from that observer's perspective, no definite fact yet exists. Conversely, once $X$ is updated with some outcome, that outcome becomes a fact relative to that observer, even if for another observer the outcome remains unrealized. Thus, the model does away with the need for a universal, absolute wavefunction; the states recorded internally by each observer are sufficient. Practically, this suggests assigning quantum states (or classical data) to observer-system pairs rather than systems

alone—the observer's internal state $X$ thus becomes part of the system's description. What the minimal observer model adds is an operational criterion for observer-relativity: an event or state is observer-relative if and only if it is recorded in (or can be inferred from) the observer's internal state $X$ as a result of an interaction crossing boundary $\mathcal{B}$.

#### THE OBSERVER'S INTERNAL STRUCTURE $(X, Y, Z)$ IN THE RQM CONTEXT

Another novel contribution of our model is that it explicitly distinguishes roles within an observer. Traditional RQM treats observers as monolithic entities relative to which states are defined. Our model introduces internal structure: $X$ (memory), $Y$ (sensor inputs), and $Z$ (actions). For instance, the input space $Y$ precisely specifies which environmental interactions the observer detects, while the internal state space $X$ explicitly limits the observer's informational capacity, reflecting the partial and coarse-grained nature of observed facts. This fine-grained structuring clarifies RQM's implicit assumption that observers record partial information. For example, an observer device with just two internal states ("event happened" vs. "didn't happen") represents the simplest possible relational scenario: relative to this minimal observer, reality reduces to binary distinctions. By contrast, richer state spaces allow more detailed descriptions, capturing progressively finer-grained relational states. The presence of action outputs $Z$ further enriches the RQM view by showing observers as active agents capable of influencing their environments. Hence, our formalism clarifies how observers interact through boundary crossings and internal updates, rigorously tracking relational facts and eliminating ambiguity about their realization and communication.

#### PHENOMENOLOGICAL SCENARIO (RQM)

Consider the well-known Wigner's friend scenario. Alice (Observer $O_A$) is inside a lab, measuring an electron's spin, while Wigner (Observer $O_W$) remains isolated outside, yet plans to measure the combined Alice-electron system later. According to RQM, the electron's spin state is definite relative to Alice but indefinite (entangled) relative to Wigner. The minimal observer model explicitly represents this scenario by assigning each observer a distinct boundary and internal state: Alice's boundary $\mathcal{B}_A$ includes her measuring apparatus, and upon observing a result (spin-up), her internal state $X_A$ is updated to encode this fact. For Wigner, whose boundary $\mathcal{B}_W$ has not yet interacted with Alice's lab, no input has entered his state space $X_W$, so from his perspective, no definite outcome has occurred. Only after Wigner interacts across his own boundary $\mathcal{B}_W$ (by observing Alice's record) does the spin become a definite fact relative to him. This provides a concrete instantiation of RQM's central thesis that facts emerge relationally upon interactions. Our model, thus, operationalizes and concretely tracks how and when relational states arise—adding significant operational clarity and formal rigor to the philosophical and conceptual assertions of RQM.

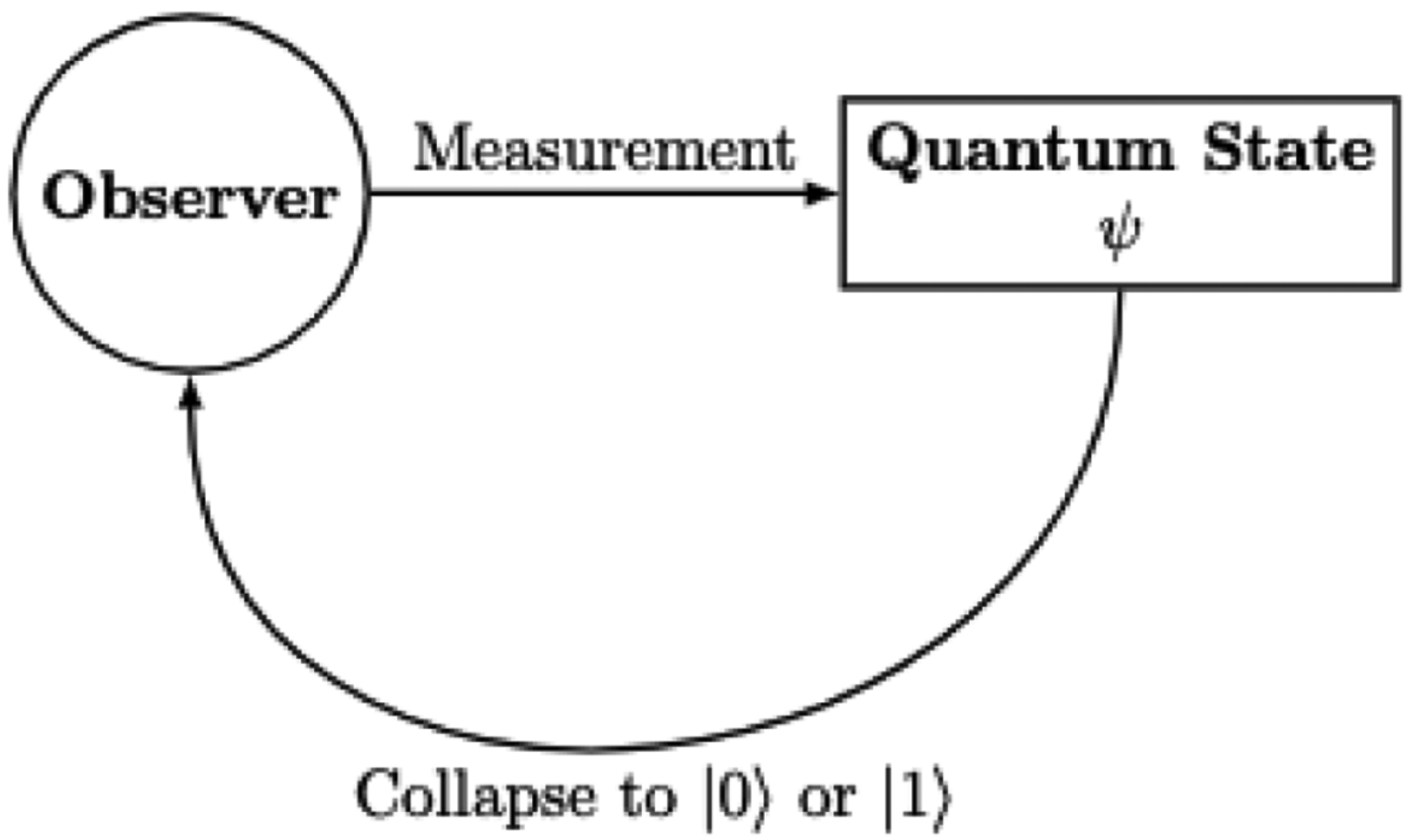

*Figure 4.9* Quantum observer measuring a quantum state.

#### *4.10.2.3 Copenhagen interpretation: A concrete classical–quantum cut via X and $\mathcal{B}$*

**THE COPENHAGEN INTERPRETATION**

The Copenhagen interpretation emphasizes the fundamental division—the *Heisenberg cut*—between the quantum system, described by a wavefunction, and the classical measuring apparatus, that records outcomes. Traditionally, the precise location of this cut is flexible, provided the apparatus and observer remain classically describable while the measured system is quantum. Measurement collapses the wavefunction, producing definite classical outcomes (Fig. 4.9), but Copenhagen interpretations leave the exact nature and placement of this boundary unclear. Debates persist regarding whether the cut involves conscious observers or macroscopic apparatuses, resulting in well-known puzzles such as Schrödinger's cat and Wigner's friend scenarios. The minimal observer model clarifies this issue by explicitly embedding the classical–quantum cut within a formal structure of the observer itself.

### THE MINIMAL OBSERVER AS THE CUT

In our framework, an observer explicitly embodies the classical apparatus along with its information-storage structure. The internal state space $X$ is defined to be classical by design, consisting of stable states (memory bits or pointer positions). The boundary $\mathcal{B}$ clearly delineates the quantum-classical interface: everything external to $\mathcal{B}$ remains quantum, while internal states in $X$ are classical records. Measurement thus involves two clear stages: (1) a quantum interaction at boundary $\mathcal{B}$ produces input $y \in Y$ (e.g., photon hitting a photographic plate), and (2) the internal state $X$ is updated to a definite outcome state via the transition function $f : X \times Y \to X$. Before measurement, the system may remain in superposition from the observer's perspective; afterward, $X$ holds a single definite classical outcome. Thus, our model provides a mathematically rigorous criterion for the condition when quantum measurement occurs, eliminating the need for vague notions of wavefunction collapse. The Heisenberg cut's position becomes explicitly adjustable through placement of $\mathcal{B}$—from immediate detector boundary to encompassing an entire laboratory—offering a quantitative foundation for Copenhagen's traditionally qualitative statements.

### EMERGENCE OF CLASSICAL RECORDS

The minimal observer model further elucidates how quantum events yield classical records. A classical record in Copenhagen terms is an unambiguously readable outcome, such as a pointer reading or detector click. In our framework, classical records correspond precisely to stable internal states $X$. Before measurement, the observer's internal state is undetermined; measurement interactions update $X$ to a stable classical state. Once recorded, this classical state persists, consistent with classical robustness ensured in experiments by decoherence and amplification. Hence, the quantum-to-classical transition explicitly corresponds to boundary-crossing inputs updating internal observer states. The model thus provides a transparent, physically grounded account of quantum state collapse without invoking consciousness or special collapse rules—only finite internal storage and boundary-based interactions.

### PHENOMENOLOGICAL SCENARIO (COPENHAGEN)

Consider a Stern–Gerlach experiment: a beam of silver atoms (quantum spins) passes through an apparatus, splitting into spin-up or spin-down components. According to Copenhagen, the atom remains in a superposition until striking a detector. Our minimal observer model clarifies this scenario explicitly. The detector plus its readout electronics constitutes an observer $O$, with boundary $\mathcal{B}$ defined at the detector's interaction surface. The observer's internal state space $X$ initially encodes a neutral "ready" state. When an atom with unknown spin interacts at the detector surface (boundary $\mathcal{B}$), an input $y \in Y$ crosses into the detector, triggering $f$ to update $X$ into a stable, definite state—either "spin-up recorded" or "spin-down recorded." This clearly defines when the quantum system transitions into a classical record. Prior to interaction, the spin state is indefinite within observer $O$; post-interaction, $X$ explicitly encodes a single classical result. Thus, the minimal observer model operationally identifies the exact moment and mechanism of quantum collapse into classical reality, addressing longstanding ambiguities within Copenhagen interpretations and providing new experimental criteria for the quantum-classical boundary.

### *4.10.3 Intrinsic space of observers and background-independent frameworks*

#### *4.10.3.1 Concept of an intrinsic observer space*

Intrinsic observer space refers to treating the set of all possible observers as a space with its own structure, rather than assuming a single "God's eye" view. In physics, one example is the manifold of all observer states in spacetime. For instance, in general relativity one can consider the 7-dimensional manifold consisting of every possible future-timelike unit tangent vector at every point—essentially "all possible observers" (each point in this observer-space corresponds to an observer with a location and velocity) [62, 63]. The goal is to describe physics from within this space of observers, using only relationships between observers, without appealing to an outside absolute frame. This is an intrinsic approach because the description is given in terms of quantities an observer can define internally (their own clock, ruler, measurement records, etc.), rather than coordinates defined by an external backdrop.

In an intrinsic observer-space formalism, one often introduces equivalence classes of observers and observer-dependent coordinates: two observers might be regarded as "the same" in some intrinsic sense if there is a transformation mapping one's observations to the other's. There is a philosophical subtlety here. One view ("view from nowhere") treats different observer frames as mere perspectives on an underlying invariant reality, identifying them via equivalence classes. Another view ("view from everywhere") holds that each observer's frame defines its own reality without needing to be quotiented out [64]. In practice, most physical theories define an observer-space along with transformations relating observers, so that one can either quotient out those differences (to find invariant physics) or consider each perspective as fundamental. Below we formalize these ideas by defining observer equivalence and coordinate structures, and then discuss how going background-independent—not only in spacetime but in the space of observers itself—leads to new formulations in category theory, differential geometry, and physics.

#### *4.10.3.2 Equivalence classes of observers*

**Observer versus frame**: In the relativistic portions of this section we momentarily use "observer'' in the *kinematical* sense of a time-like world-line equipped with a local orthonormal frame (or tetrad). That frame can be dragged by a Lorentz transformation in special relativity, or by a spacetime diffeomorphism in general relativity; the full, feedback–loop observer of Section 4.1 is therefore represented by an *orbit* (gauge class) of such frames rather than a single coordinate chart.

Observer equivalence means that under certain transformations, two observers are considered physically or functionally the same. In relativity, for example, the relativity principle states that all inertial frames are fundamentally equivalent—no preferred inertial frame exists. This can be phrased as: "all reference systems are equivalent," reflecting a democratic principle that physics should not single out one observer over another [65]. Any two inertial observers moving at constant velocity relative to each other are related by a Lorentz transformation; they form a single equivalence class under these symmetry transformations. More generally, in general relativity any two observers (inertial or not) can be related by a diffeomorphism (a smooth coordinate transformation) acting on the spacetime manifold. If a diffeomorphism maps one observer's worldline and measurements to another's, the two are considered the same physical situation due to general covariance (diffeomorphism symmetry). In gauge-theoretic language, choosing a different observer or reference frame is often akin to a gauge transformation—a change of description that does not alter physical content [62, 64]. Thus, an equivalence class of observers can be defined by the symmetry transformations of the theory: all observers related by a valid change-of-frame (Poincaré transformations in special relativity, or arbitrary diffeomorphisms in general relativity) belong to the same class, representing one "physical situation" viewed from different perspectives.

In more abstract terms, one can formalize observers and their transformations using groupoids or categories. We may define a category of observers where each object is an observer (or an observer's coordinate system) and each morphism is a change-of-observer transformation (such as a coordinate change or Lorentz boost). This category is naturally a groupoid (every transformation is invertible), encapsulating the idea that all observers are on equal footing and can transition into one another's perspective [65]. Within this framework, an isomorphism (invertible morphism) between two observer objects indicates that they are effectively the same observer in a relational sense. Put differently, an isomorphism is a dictionary translating one observer's internal descriptions into another's. Using this idea, one can rigorously define an equivalence relation: $O_1 \sim O_2$ if and only if there exists a bijective structure-preserving mapping (an isomorphism) between observer $O_1$ and $O_2$. It can be shown that this $\sim$ indeed satisfies reflexivity, symmetry, and transitivity (i.e. it is an equivalence relation). In practical terms, $O_1 \sim O_2$ means the two observers differ only by a relabeling of their relevant structures (such as coordinate labels or internal states) while preserving all relationships—they are "the same observer described in two different languages."

For example, in a formal observer-as-system model (discussed more in the computational context below), one theorem states that two observer systems are equivalent if there is a bijection between their internal state sets, input channels, and output channels that makes their transition and observation dynamics coincide; this bijective homomorphism defines an observer isomorphism. In other words, two observers are "equivalent" if they differ only by relabeling their internal states, inputs, and outputs in a way that preserves the structure of their feedback and state-transition dynamics. Thus we obtain equivalence classes of observers as sets of all observers isomorphic to each other.

By factoring out (i.e. identifying) observers related by these transformations, physics focuses on the invariant content. However, one may also study the space of inequivalent observers as a manifold or set of distinct vantage points. Clearly defining the equivalence relation is crucial—whether it's "same physical trajectory up to re-labeling," "same dynamics up to relabeling of states," or some other criterion depending on context. Moreover, one must decide whether to adopt the view that only the equivalence class has objective meaning (the view-from-nowhere stance) or that each equivalence class member may be treated as having its own reality unless/until related to another (the view-from-everywhere stance) [64].

#### *4.10.3.3 Observer-dependent coordinates and frames*

Each observer in the intrinsic space typically comes equipped with their own coordinate system or frame of reference. That is, an observer defines how to measure space and time (or other quantities) relative to themselves. Two different observers generally have different coordinate descriptions of the same events. Formalizing this, one can assign to each observer a coordinate chart or a basis of measurement. For instance, an observer in spacetime has a natural choice of time coordinate (their proper time along their worldline) and space coordinates (defined by some simultaneity convention, like projecting events orthogonally to the observer's worldline). This leads to different splittings of spacetime into "space" and "time" for different observers.

A simple example: in special relativity, two inertial observers moving relative to each other have time axes that mix into each other's space axes—what one calls "now" is a mixture of "now" and "later" for the other, due to the relativity of simultaneity. In general relativity or accelerating frames, the differences are even more pronounced. A notable case is the Rindler observer (uniformly accelerated) vs. an inertial Minkowski observer: each has a different foliation of spacetime into space+time. The accelerated Rindler observer uses Rindler coordinates, in which their constant-time slices are hyperbolas that an inertial observer would describe as accelerated trajectories. This difference in "observer's space" has concrete physical effects—each observer literally perceives a different version of space and time. For example, the Unruh effect can be interpreted as arising from this difference: an accelerated observer (Rindler frame) sees a thermal bath of particles, while an inertial observer sees vacuum, because the very notion of what constitutes a "particle" (or even a vacuum state) depends on the observer's space–time splitting [66]. In cosmology as well, an "observer's space" can be defined at each moment for an arbitrary worldline via a chosen simultaneity prescription; this space generally differs from any globally preferred slicing (e.g., an inertial observer in an expanding universe defines space in a way that is neither homogeneous nor isotropic, even if the universe has a homogeneous slicing) [66]. All these examples underscore that coordinate structures distinguishing observers are fundamental—one must carefully specify which observer's frame is being used when assigning values to measurements.

Mathematically, one way to encode observer-dependent frames is with the frame bundle or observer bundle on spacetime. The bundle of orthonormal frames at each spacetime point includes all possible orientations of an observer's axes at that point. Picking a specific frame (a basis of time+space directions) is like choosing an observer at that event. In differential-geometric terms, the space of observers can be modeled as this bundle: each point in observer-space might be labeled by $(x, u)$, where $x$ is a spacetime point and $u$ is a unit timelike 4-velocity at $x$ (the observer's local velocity or time-direction). One can introduce coordinates on this observer-space: for example, coordinates $(t, x, y, z, v_x, v_y, v_z)$ could label an observer's spacetime position $(t, x, y, z)$ and their 3-velocity components $(v_x, v_y, v_z)$ (or analogous parameters like rapidity and orientation angles). Such coordinates on observer-space describe how observers differ by location and state of motion. The transformations between observers (e.g., Lorentz boosts, rotations, or general coordinate changes) will act on these coordinates. By studying the observer-space as an entity, we can attribute geometric structures to it (such as a metric or connection defined on $O$). The key idea is that an observer's coordinate system is not global and absolute; it is attached to the observer. When comparing two observers, we either transform one's coordinates into the other's or work in the larger space containing both and map between their coordinate patches.

One concrete example of defining observer-dependent coordinates is the construction of a canonical reference frame for any given observer in curved spacetime. Lachièze-Rey [66] provides a prescription to foliate spacetime according to an arbitrary observer's notion of simultaneity. This yields a unique slicing (and thus a set of spatial coordinates at each instant) corresponding to that particular observer, even if the observer is accelerating or in a general cosmological model. The result is that different observers have different "slices" of space—different 3D surfaces that they consider to be happening "now"—and thus they decompose spacetime differently. The Langevin observer (in the twin paradox) or a rotating observer, for example, will assign a different geometry to space (non-Euclidean, etc.) than an inertial one [66]. All these constructions are ways of giving each observer an intrinsic coordinate map. Such a coordinate structure is crucial for observer-dependent physics (like defining particle horizons or energy as measured by that observer).

In summary, the intrinsic space of observers comes with a myriad of possible coordinate choices, one per observer. The relations among these coordinate systems (through transformations) are what connect one observer's measurements to another's. This viewpoint emphasizes that many quantities (length, time interval, even the vacuum state of a field) are not absolute but depend on the observer's frame. By formalizing an observer-space, we keep track of these dependencies systematically and can define what it means for a quantity to be covariant or invariant under change of observer. An invariant would be something all observers agree on or that transforms trivially (e.g., the spacetime interval, or an abstract action integral), whereas a covariant quantity has a well-defined transformation law that allows any observer to calculate what another would see.

#### *4.10.3.4 Extending background independence to observer space*

Background independence traditionally means a theory does not presume a fixed spacetime structure—the spacetime geometry is dynamical or undetermined, and only relationships (e.g., causal or metric relations) that satisfy the equations have physical meaning. General relativity is the prime example: there is no fixed background metric; the geometry (metric field) is a part of the solution. Extending this idea beyond spacetime to the space of observers means the theory also does not assume a preferred or fixed observer/frame of reference. In other words, not only is spacetime relative, but the arena of all observational viewpoints is itself treated without prior structure. All observers are on equal footing, and the laws of physics must be formulated without secretly choosing a special observer or coordinate system in advance. This is essentially the principle of relativity and general covariance taken to the next level—it suggests that the space of observers is taken as fundamental, and spacetime (with its events and distances) might even be secondary or emergent from relations among observers [62].

One concrete realization of this idea comes from Cartan geometry and frame bundle techniques. In a fully background-independent view, one can start with the space of observers as a primary object and later extract spacetime as a derived concept. For example, consider again the observer manifold $O$ (all events + a unit timelike direction at each event). On this 7D manifold, we can define fields and a Cartan connection such that no *a priori* spacetime metric is needed—the only input is the local symmetry group (the Lorentz group) which acts on the fibers. Gielen and Wise propose exactly this: "taking observer space as fundamental" and formulating gravity in terms of an observer-space geometry [62]. In their approach, spacetime can be recovered as a quotient of the observer space if certain integrability conditions hold (essentially, if a global 4D slice through the 7D space can be consistently defined). If those conditions fail, it suggests a scenario in which an absolute spacetime cannot be stitched together—instead, physics might only be definable in terms of overlapping observer perspectives, making spacetime an observer-dependent, relative concept [62]. This radical possibility is a form of fully relational ontology: the fundamental description is a network of observer states and their relations, with no single universal spacetime backdrop for all events.

Even without going to such extremes, insisting on observer-background independence means our formalism should be covariant under change of observer. In practical terms, any statement or equation should be valid in any frame—or transform appropriately between frames—without relying on a fixed background frame. Category theory provides a natural language for this: one can require that the theory is formulated as functorial or natural with respect to the category of observers. For instance, in a category-theoretic formulation of quantum physics or gravity, one proposal is to assign to each observer their own state space (Hilbert space) and then relate these via functors. Crane suggests that a "state for quantum gravity" could be described as a functor from the category of observers to the category of vector spaces (Hilbert spaces) [67]. What this means is that each observer (object in the category) gets a vector space of states, and an observational transition (morphism) between observers (representing a change of reference frame or perspective) induces a linear map between their state spaces. Physical laws would then be invariant under such changes if they arise as natural transformations between these functors (so that the state assignments to each observer are consistent with evolving or transforming the system). This categorical construction embodies background independence at the level of observers: there is no single "preferred" state space or single observer's coordinates in which the theory must be formulated. Instead, the principle of relativity is built in—any observer's description can be transformed to any other's systematically.

Another way to enforce no preferred observer is to treat the full groupoid of observers as the arena for physics, instead of a single spacetime. As mentioned, one can view all frames as connected by morphisms. The laws of physics should be expressible in a way that is invariant under moving along these morphisms.

In effect, this is like saying the fundamental formulation is done "up on $O$" (the observer space) rather than down on $M$ (spacetime) with a chosen frame. When done properly, this yields all the same physics but with a manifest guarantee of symmetry. A simple analogy is how one formulates Maxwell's equations or other laws in tensor form—by writing them in a covariant form, one shows they hold in any coordinate system (observer). Here we elevate that idea: we formulate the entire theoretical framework on a structure that does not bias any particular observer or coordinate system. Only relations between observers (like relative velocities, or intersection events) might enter.

In practical terms, achieving full background independence including observers might require additional constraints or symmetry principles. For example, requiring that no global structure (like a global time coordinate or global inertial frame) appears, forces the introduction of fields or connections that compensate for shifting observer perspective. The observer-space approach introduces something like an "internal observer group" gauge symmetry—shifting an observer's 4-velocity direction (i.e. moving along the fiber at fixed spacetime point) is a symmetry operation. Demanding invariance under this symmetry can lead to new conservation laws or conditions (just as gauge invariance leads to conserved currents via Noether's theorem). In short, extending background independence to observers means that the theory's degrees of freedom and constraints should be described in a way that does not require fixing an observer. This is a natural generalization of general covariance: not only are coordinates unphysical but even the choice of an observing frame is unphysical until an interaction (measurement or communication) relates two observers.

The payoff of this extension is conceptual clarity and universality. It becomes clear which aspects of a theory are genuinely invariant and which are convention-dependent. It also helps bridge physics with information theory and computation, where the role of the observer (or agent) is crucial. By not pinning down an observer, we keep the theory general enough to apply, say, to any agent (human, machine, or particle) that could be making observations. The theory then has to supply rules for how different agents' accounts compare—which is precisely the role of the transformations in observer space or the natural transformations in the categorical approach. This fully relational perspective is at the heart of many modern discussions in quantum gravity, quantum foundations, and even philosophy of science, where one tries to eliminate any lingering absolute structure, including the abstract "observing subject," and replace it with a web of relations.

#### *4.10.3.5 Category-theoretic perspectives on observers*

Category theory offers a high-level, structural way to describe observers and their interrelations. We already touched on the category of observers idea, where observers are objects and transformations between observers (changes of frame or perspective) are morphisms. Because any observer should in principle be able to transform to any other (given the appropriate coordinate transformation or data translation), this category is typically a groupoid (every morphism is invertible). Formulating physics in this language makes symmetries and equivalences explicit. For example, the collection of all inertial frames in special relativity can be seen as the objects of a category, with a morphism for each Lorentz transformation mapping one frame to another. That category has the structure of the Poincaré group action (essentially a one-object category if we identify all inertial frames as instances of the same abstract frame, or a groupoid if we treat each frame as a separate object). The advantage of the categorical view is that one can then impose conditions like functoriality or naturality to ensure physics is consistent across different observers.

A striking use of category theory in observer-related physics is in quantum gravity and quantum foundations. The work of Crane (1993) and others envisioned a scenario where each observer has their own Hilbert space of quantum states, and physics is a kind of many-object generalization of quantum mechanics. In this approach, consistency between observers is maintained by categorical structures. Specifically, Crane described that "a state for quantum gravity is given by a functor from the category of observers to the category of vector spaces" [67]. In plain terms, this assigns to each observer a vector space (typically a Hilbert space) such that when you have a morphism (an observation change or reference-frame transformation) from observer $A$ to observer $B$, the functor provides a linear map between $A$'s vector space and $B$'s vector space. The physical state of the universe would then not be a single vector, but rather the entire functor—which consistently assigns state vectors to each observer and ensures that if two observers are related by a transformation, their state descriptions are related by the corresponding linear map. Time evolution or other processes can be described by natural transformations (mapping functors to functors) which play the role of dynamics in this schema. This categorical formalism is closely related to the idea of a topological quantum field theory (TQFT), where a state is associated to boundaries (observers can be thought of as "boundaries" between observed system and observer) and consistency conditions must hold for gluing boundaries (analogous to communicating or transforming between observers).

Another category-theoretic approach is the groupoid model of relativity. As mentioned, instead of focusing only on a symmetry group (which usually has one object/state and many automorphisms), one considers the category of all observers with morphisms as allowed transformations. This can handle cases that a single group cannot (e.g., when only certain observers can directly transform to each other, or when composition of transformations has path-dependence as in general relativity's gravitational holonomies). In a groupoid, each object has its own little group of automorphisms (its symmetry group of leaving that observer invariant), and the whole structure can encode both symmetry and the equivalence relation between different frames. Oziewicz proposed to "formulate the physics of relativity in terms of the groupoid category of observers, keeping strictly the most democratic interpretation of the Relativity Principle that all reference systems are equivalent" [65]. This means rather than starting with a fixed space and one group acting on it, one starts with the many-object category where each object is a reference frame and each morphism is a change of frame, and one builds the theory there. Doing so can reveal hidden assumptions of the usual approach—for instance, the breakdown of a single-group picture when considering non-inertial observers or gravity can be naturally accommodated by a groupoid (since accelerating frames might not be related by a single global Lorentz transformation, but can be related piecewise). The category approach also generalizes to observers of different types (imagine a category that includes classical and quantum observers as different kinds of objects, and morphisms that describe interactions or translations between their descriptions).

Category theory also provides tools for hierarchies of observers. One can consider 2-categories or higher categories where morphisms between observers themselves have morphisms (think of one observer observing a pair of other observers in communication—this could be a 2-morphism in a higher category of observers). These abstractions can formalize complex scenarios like "observer $A$ watches observers $B$ and $C$ conduct an experiment." While these are mostly theoretical at this stage, they hint at a unified language for multi-observer interactions.

In less abstract terms, categories help enforce that physics is independent of the *choice* of observer by design. If the theory is formulated as a functor on the *category of observers*, then by definition it assigns equivalent data to *isomorphic* observers—so every member of an equivalence class receives the same physics. This is a powerful way to encode observer-independence: physical predictions arise as functorial assignments that do not discriminate between isomorphic viewpoints. Any quantity that is *strictly* observer-invariant will factor through this functor to the *quotient category of equivalence classes*. Importantly, that quotient is not a return to a single, objective "view from nowhere"; it is itself defined *relationally*, via the mappings generated by observers (a stance fully aligned with second-order cybernetics [15]). By contrast, quantities that remain observer-relative live entirely in the original category and transform non-trivially along its morphisms.

To summarize, category-theoretic perspectives treat observers as fundamental objects and changes of perspective as fundamental morphisms. They naturally encode equivalence (via isomorphisms) and can enforce that the theory treats all observers without favoritism (via functorial assignments). This is a very general framework, capable of bridging physics with computer science and logic (where "observers" could be seen as contexts or processes, and one uses categorical semantics to relate them). The price is a high level of abstraction, but the payoff is unifying disparate ideas (symmetry, relativity principle, reference frames, information transfer) under one mathematical roof.

#### *4.10.3.6 Differential-geometric observer space structures*

Differential geometry provides another powerful framework for formalizing an intrinsic space of observers. Here, one treats the collection of observers as a manifold or fiber bundle and endows it with geometric structures. A clear example is again the set $O$ of all future-directed unit timelike vectors in a given spacetime (assuming a Lorentzian manifold $M$). $O$ can be thought of as the unit timelike tangent bundle of $M$—it is a fiber bundle over spacetime $M$, where each fiber (at a point $x \in M$) is the hyperboloid of unit timelike vectors (the possible 4-velocity directions for an observer at $x$). For 4-dimensional spacetime, $O$ is 7-dimensional (4 for position + 3 for the velocity direction). We can call $O$ the observer manifold. Now, any field on $O$ that is appropriately invariant can represent a physical quantity measured by observers. For instance, a function on $O$ could assign to each possible observer a value (like "the temperature that observer measures")—different values for different states of motion if the effect is observer-dependent (as with Unruh radiation). More powerfully, one can formulate dynamics directly on $O$. Gielen and Wise demonstrated that one can reformulate general relativity in terms of a Cartan geometry on observer space [62]. In their formulation, the usual Einstein field equations can be derived from curvature conditions and fields on $O$, and spacetime itself emerges as a derived concept (as an equivalence class or quotient of observer trajectories). The geometric idea is that $O$ comes naturally equipped with two distributions (roughly: horizontal directions correspond to moving an observer in spacetime and vertical directions correspond to changing an observer's velocity at the same spacetime point). A Cartan connection on $O$ can encode both the spacetime curvature and how local reference frames rotate or accelerate. The equivalence principle (local Lorentz symmetry) is built in by the fact that the structure group on $O$ is the Lorentz group, which acts on the fibers (different velocity directions at one point are related by Lorentz transformations). By using Cartan geometry, one ensures that at each point of $O$ the geometry looks like a "model geometry" (Minkowski space for the horizontal part, and a velocity-space model for the vertical part).

One major benefit of a differential-geometric observer space formalism is that it can handle observer-dependent effects in a smooth, quantitative way. Notions like "spatial vs temporal direction" become geometric: at each observer-state in $O$, one can identify the subspace of directions that correspond to that observer's spatial axes (those directions perpendicular to the observer's 4-velocity) and the time direction (along the 4-velocity). In the Cartan setup, these come from splitting the tangent space of $O$ using the observer's velocity field. Geometric structures on $O$ can encode things like an observer's proper time (a natural time coordinate along the observer's worldline in $O$), gravitational fields (which might appear as curvature or torsion in the connection on $O$), and fictitious forces in non-inertial frames (which appear as real geometric effects in an accelerating observer's bundle). In this picture, a specific observer is represented as a curve in $O$ (e.g., an observer moving through spacetime traces out a path through different points of $O$, since their position and velocity may change). The physics along that curve is that observer's experience. But because we have the whole manifold $O$ that includes all other observers, we can relate different observers by geometric relations in $O$. For instance, two observers might come into contact (literally meet at an event): this is represented by two curves in $O$ intersecting at a point (meaning they share the same $x$ and $u$ at that instant). Or an observer accelerating is moving vertically in $O$ (changing $u$ while staying at roughly the same $x$, in an instantaneous sense). The geometry of $O$ might tell us, for example, how an accelerating observer's spatial slice tilts and how their notion of simultaneity shifts—information encoded in the connection on $O$.

Importantly, this approach yields insight into background independence. If $O$ is fundamental, we do not assume spacetime $M$ exists as a separate stage; it can be constructed from $O$ by identifying all observer-states that share the same event regardless of velocity (that quotient gives back spacetime $M$ if it exists). The conditions for this to work are essentially that certain fields on $O$ (like an observer congruence field) are integrable. When those conditions are met, the standard spacetime picture emerges smoothly from the observer picture. When they are not, it suggests something like a "foamy" situation where different observers' views cannot be stitched into a single manifold—possibly an avenue to understand quantum gravitational scenarios where classical spacetime breaks down. But even aside from that extreme, working on $O$ has practical advantages. In $O$, one can separate "absolute" properties from observer-relative ones. For example, a tensor field on spacetime, when pulled back to $O$, can be split into parts seen by a given observer (like splitting an electromagnetic field into electric and magnetic fields depends on the observer's velocity). On $O$, that splitting is just evaluating the field with the additional data of $u$. Maxwell's equations can be pulled back to $O$ and yield a set of equations that explicitly show how different observers see electric/magnetic fields mix. In gravitational dynamics, the Hamiltonian (canonical) formulation chooses an observer foliation; working on $O$ lets one derive the Hamiltonian constraints without ever explicitly choosing a foliation—instead, an "observer field" (a choice of a rep-resentative observer at each spacetime point) can be considered a gauge fixing that $\mathcal{O}$ allows us to handle flexibly [62].

Another differential-geometric insight is how measurement invariants appear in observer space. In general relativity, an "observable" must be invariant under diffeomorphisms (since coordinates are arbitrary). In observer-space terms, an observable might be a function on $\mathcal{O}$ that is invariant under the local Lorentz transformation on the fibers (because changing the inertial axes of a given observer shouldn't change a scalar physical quantity). Thus true invariants live on $\mathcal{O}$ but do not depend on the $u$ aspect—only on the spacetime event (like proper scalar curvature at a point). Those correspond to usual scalar invariants in spacetime. However, one can also consider observer-dependent observables—quantities that *do* depend on $u$, i.e. on the observer's state of motion. These are not invariants of the full diffeomorphism + Lorentz gauge, but they are well-defined as functions on $\mathcal{O}$. An example is the energy density of a field as measured by an observer with 4-velocity $u$. This is not an invariant scalar on spacetime (it depends on the observer), but it is a well-defined scalar on $\mathcal{O}$. By working on the observer manifold, we can talk about such quantities legitimately and track how they change as one moves in $\mathcal{O}$ (i.e. as the observer changes). This can be useful in relativistic statistical mechanics or black hole thermodynamics, where one wants to compare what different families of observers see.

In summary, the differential-geometric approach builds an intrinsic coordinate system on the space of observers itself. It treats observer transformations as fundamental symmetries (a kind of extended gauge symmetry that includes reference-frame changes). This allows us to express laws of physics in a manifestly observer-covariant way. Ultimately, such formalisms impact how we think of space, time, and measurement: they blur the line between what is "physical (spacetime) geometry" and what is "perspective." All frames live in one big space, and a given frame's coordinates are just one patch on this observer manifold. This is an explicit way of enforcing no preferred frame—the geometry doesn't care which observer you label as origin because any point in $\mathcal{O}$ is just as good as any other for describing physics. It also provides new tools to analyze physical problems by lifting them to $\mathcal{O}$, solving symmetrically, and then projecting results back down to particular observers.

#### *4.10.3.7 Observer-dependent physics and relational measurements*

Physical theories increasingly recognize that what is measured or observed can depend on the state of the observer. We have already seen examples in relativity (time dilation, length contraction, simultaneity shifts, particle detection differences in the Unruh effect) where different observers experience different values or even different qualitative phenomena. By formalizing the space of observers, we get a handle on how to transform measurements from one observer to another and what structures are invariant versus what are observer-dependent. In classical physics, these transformations are given by kinematic symmetry groups (Galilean or Lorentz transformations). In modern physics, we encounter observer-dependence in broader contexts, for example:

**Quantum measurements:** In quantum mechanics, the result of a measurement can depend on the "context"—which is often tied to the observer's experimental setup or frame. Different observers (especially in thought experiments like Wigner's friend) might not even agree on what has been measured or the state of a system. Relational quantum mechanics (Rovelli) posits that the quantum state is not absolute but is relative to each observing system. This is analogous to how in relativity an event's time coordinate is observer-dependent; here, the outcome (or state assignment) is observer-dependent, and only when two observers exchange information and correlate their records do they find a consistent story. An observer-space for quantum contexts could formalize this by letting each observer have their own space of possible quantum states of the world, and "bridging maps" when observers interact and compare notes. Such ideas are under development, often using category theory or extended Hilbert space formalisms.

**Thermodynamics and horizons:** As mentioned, an accelerated observer perceives a horizon; for example, an observer free-falling into a black hole and one hovering just outside it register drastically different phenomena, even though they are both describing the same underlying physics—namely, the covariant, horizon-free field equations. Temperature and entropy can be observer-dependent. The entropy associated with a horizon (like black hole entropy or de Sitter horizon entropy) might be seen as an observer-dependent count of inaccessible information. In an observer-space picture, one might label points not just by location and velocity but also by region of spacetime accessible to that observer (horizons create a partition of what can be observed). Then laws like the second law of thermodynamics might hold in a form that depends on that partition. Cosmological observations too depend on the observer's worldline (our current observations of the universe are from one very specific vantage point). When we talk about the universe's properties, we often implicitly mean "as seen by comoving observers" or "as would be seen by an ideal inertial observer at rest with respect to the CMB." Transforming to another observer (say moving at 0.9c relative to the CMB rest frame) would complicate those properties (the CMB would be highly anisotropic, etc.). Formally including observers in the model helps make those dependencies explicit and thus clarifies which statements are invariant.

**Gauge and symmetry breaking:** Sometimes choosing an observer can be like choosing a gauge in field theory. For instance, in the Higgs mechanism, one typically works in unitary gauge to interpret the physics, which is analogous to working in the rest frame of the Higgs field's "observer." If one chooses a different gauge, the interpretation changes. In gravitational physics, choosing a particular time slicing (observer family) can break time-translation symmetry that might otherwise be present. Thus, observer choices can effectively break symmetries that the underlying equations have, leading to different conserved quantities or lack thereof. Only by checking invariant structures (like energy measured at infinity, etc.) can one get observer-independent conclusions.

By extending our framework to include observers, we also get a clearer picture of what an "observation" fundamentally is. In an observer-centric view, an observation is an event that involves both an object system and an observer system, resulting in a correlation between them. For example, a measurement in quantum mechanics entangles the apparatus (observer) with the measured system; in classical terms, a measurement imprints information about the system onto the observer's state (like a meter reading). In a relational view, the basic ingredients are triadic: an observer, an observed phenomenon, and the interaction linking them. If we imagine the space of all possible such interactions, that itself might be structured (one could use category theory here, too, with interactions as morphisms between observer and system). The feedback loops we discuss next build on this idea that observation is not one-way—the observer can influence the system as well.

Ultimately, making physics observer-dependent (in the formalism) doesn't mean giving up objectivity, but rather refining what objectivity means. It means that a statement is objective if it is formulated in the language of the observer space and does not actually depend on which observer-state we pick (or if it does, we know exactly how to translate between them). It's similar to how in general relativity an "objective" statement is one that is tensorial (covariant)—you can write it down in any coordinate system and it's true in all. Here, an objective statement might be one that, say, all observers agree upon when they compare (like a properly invariant scalar), or a relationship that holds between any two observers' measurements when transformed appropriately. By contrast, something like "observer $O$ sees a particle with energy $E$" is not objective by itself; but "observer $O$ sees a particle with energy $E$ *and* observer $O'$ (moving at $X$ relative to $O$) sees it with energy $E'$, related by the Lorentz factor" is a complete, transformable statement. Observer-space formalisms strive to encode such complete relations from the start.

#### *4.10.3.8 Measurement and feedback cycles with embedded observers*

When the observer is included as an integral part of formalism, the act of measurement is no longer a passive reading of a pre-existing value—it becomes an interactive, dynamical process. Measurement can be thought of as a mapping from the observed system's state to the observer's own state (e.g., a thermometer absorbing heat and its mercury rising, encoding the temperature). In an intrinsic observer framework, one explicitly represents this mapping. For instance, in a minimal observer model (common in cybernetics and control theory), we have an observer with an internal state space $X$, receiving inputs $Y$ (sensory data) and producing outputs $Z$ (actions or signals). The observer's update rule $f : X \times Y \to X$ takes the current internal state and a new input to produce an updated state, and an output rule $g : X \to Z$ generates an output based on its state. This quintet $(X, Y, Z, f, g)$ defines a simple observing system. Measurement events in this model are inputs $y \in Y$ from the environment that cause state transitions $x \to f(x, y)$; the outcome of the measurement can be considered the pair of new state $x'$ and perhaps an output $z = g(x')$ (if the observer announces or uses the information). Crucially, the observer's state is altered by acquiring information—the observer "remembers" or reflects the measurement.

This naturally leads to a feedback loop when we allow the observer to not only sense but also act. The observer's output $z$ might influence the environment or the system being observed. In engineering terms, the observer (or controller) might then affect the next input it receives. Thus, we get a closed-loop system: environment state → sensor input to observer → observer state update → observer output action → environment changes → new sensor input, and so on. Cybernetics has long studied such loops, emphasizing that the observer (or agent) and the environment co-evolve in response to each other. The intrinsic observer concept directly incorporates this: an observer is not an abstract entity outside the system but a subsystem engaged in a feedback cycle. The field of second-order cybernetics explicitly considers the observer observing the system and itself. Key ideas from second-order cybernetics include: (1) *Observer inclusion:* the observer is part of the feedback process, not a neutral external vantage point; (2) *Self-reference:* the observer can observe and modify its own state or rules, leading to learning or adaptation; (3) *Constructivism:* what is perceived as reality emerges through the observer's interactions and interpretations, rather than being a fixed external truth.

Bringing these ideas into physics and computation, we analyze feedback loops in measurement. For a physical example, consider a robotic sensor observing a pendulum. The robot reads the pendulum angle (input), then perhaps adjusts a motor (output) to change the pendulum's motion, maybe to stabilize it. The robot is an observer with a goal (keeping the pendulum upright). Its internal state might include an estimate of the pendulum's angle and angular velocity (an internal model). It continually updates this estimate with sensor readings and outputs motor torques. This setup can be described in the observer-space framework: each state of the robot (observer) together with a state of the pendulum (system) is a point in a combined space, and the dynamics form a closed loop. To analyze it properly, one must consider both together—the combined system has no fixed external reference, it's just two interacting parts. But one can also adopt the robot's perspective: from its "intrinsic" view, it tries to measure and control the pendulum, treating itself as the reference. If we swap out the robot for a different controller with a different internal mechanism, the outcomes differ—this is essentially a different observer in the intrinsic space, and whether it can achieve the goal or make the same measurements is an observer-dependent matter.

Given a formal observer model, we can define when two observers are equivalent in their measurement and control capabilities. Earlier, we discussed an equivalence relation based on a bijective homomorphism between two observers' state-input-output structures. That theorem essentially states that if you can relabel the internal states and I/O of observer $O_1$ to get $O_2$ such that their update ($f$) and output ($g$) functions correspond exactly under that relabeling, then the two observers are behaviorally identical—they will react to inputs and produce outputs in the same way up to renaming. In terms of measurement and feedback, this means no outside entity could tell the difference between $O_1$ and $O_2$ by interacting with them. This idea is very close to the concept of bisimulation in computer science: two systems are observationally equivalent if an external observer cannot distinguish their behaviors via any sequence of tests. Here we are considering the *observers* as the systems of interest—so we are looking at equivalence from a meta-perspective. If two observers are equivalent (isomorphic), they have the same "observational power" and the same kind of feedback dynamics. They belong to the same equivalence class in observer-space and can be treated as the same point if we quotient out those symmetries.

Considering feedback cycles also raises the question of stability and adaptation. An observer with a feedback loop might reach a fixed point or a limit cycle in its state (e.g., a thermostat will reach an equilibrium temperature reading when the room stabilizes). If we change the observer (say, make the thermostat twice as sensitive), the equilibrium might shift—but there might be an invariant (like the fact that equilibrium is when room temp equals target temp, regardless of sensitivity). Understanding these feedback invariants is a part of a generalized observer theory. In this study, for example, results are derived concerning loop efficiency and adaptation speed, which depend on the observer's structure (such as its response time) but not on arbitrary labeling. This illustrates that within the space of observers, one can define metrics or partial orders: some observers are "faster," "more complex," or "more capable" than others, in ways that are invariant under relabeling (so they are intrinsic properties of the equivalence class). For instance, an invariant might be the number of internal states or the presence of a certain feedback sub-loop. These invariants help classify observers beyond simple equivalence. In physics, one could imagine classifying observers by their acceleration (which distinguishes an inertial vs. non-inertial class) or by their field of view (horizon or no horizon), etc.

Finally, embedding the observer clarifies the measurement uncertainty and disturbance. In quantum mechanics, this is usually discussed via the uncertainty principle and back-reaction. In a fully observer-space approach, the measuring apparatus is just another physical system (another "observer") interacting with the system of interest. So one can, in principle, track how the joint system's state evolves under interaction and see the trade-off—information gained by the apparatus corresponds to something (like entanglement or disturbance) in the system. In classical terms, including the observer's dynamics can show how measurement noise and delays affect results. A laboratory measurement often involves a chain of observers: for example, a particle's position influences a detector (observer 1) that converts it to an electrical signal, which is read by a computer (observer 2), and interpreted by a scientist (observer 3). Each link is an observer relative to the previous stage. Only by considering them together can we understand the full measurement record. The intrinsic observer framework encourages thinking in this compositional way—observers observing observers, etc., which category theory handles well (via composition of morphisms).

In computational contexts, these feedback considerations are concrete. In software or AI, an agent observing an environment and adjusting to it can be modeled by the same kind of state-machine observer described above. The concept of observational equivalence in computer science (two programs are equivalent if no test can distinguish them) is directly analogous to the observer-isomorphism idea. In fact, one can think of an algorithm as an observer of the input data: two algorithms are observationally equivalent if for every input (stimulus) they produce the same output (response)—this is essentially the idea of two functions being extensionally equal, or two state machines being bisimilar. The formalism we discussed thus bridges to computer science: an observer is basically an abstract machine processing inputs to outputs. The earlier theorem establishing an equivalence relation on observers by homomorphism is very much a computing notion (an isomorphism of state machines) cast in our generalized observer language. This underscores that the space of observers can include not just physical observers (people, particles, detectors) but also computational observers (algorithms, robots, AI agents), and the same formal ideas apply.

By treating observers and their interactions as first-class entities, we gain a unified perspective on measurement, feedback, and the relational nature of observation across disciplines. Physics gains a language to incorporate the agent who is observing, and computer science gains physical insight (e.g., any computation can be seen as an interaction in some physical substrate observed by some entity). Observer-space formalisms thus impact how we understand knowledge and information: knowledge is no longer an abstract absolute; it is something held by an observer, and information is what is communicated from one observer to another. The feedback loop viewpoint also emphasizes learning and adaptation, which are crucial in fields like robotics and even in evolutionary contexts (organisms as observers of their environment, adapting via feedback). All these perspectives indicate that defining an intrinsic space of observers and insisting on background-independent, equivalence-respecting structures provides a powerful framework. It forces us to carefully distinguish what is observer-specific from what is truly universal, and it provides mathematical tools (from group theory, category theory, and differential geometry) to navigate between perspectives. In doing so, it enriches our understanding of measurement processes, ensures consistency across different viewpoints, and potentially helps reconcile differences between how computations/observations occur in different domains (quantum vs. classical, physical vs. virtual). Ultimately, it highlights that observation is an active, context-dependent process—one that can be formalized and studied on its own terms, rather than always being externalized or ignored.

### 4.10.4 Mathematical formalization and theorems

Having presented the conceptual framework of *intrinsic observer space*, category-theoretic perspectives, and background independence in the previous sections, we now turn to a precise algebraic formalism. Our goal is to define how one observer may be mapped to another in a manner preserving the structure of state transitions and outputs, thereby showing a formal equivalence that mirrors the isomorphisms discussed earlier. We also introduce quantitative measures (complexity and adaptation speed) that capture how observers process and respond to inputs in a feedback loop.

#### 4.10.4.1 Observer equivalence and invariants

*Remark 4.1*—**On commutativity and "homomorphisms":** Strictly speaking, the mappings introduced below capture the requirement that the relevant transition diagrams *commute*: for any internal state $x$ and input $y$, the map $\phi_X$ must intertwine with $f$ to preserve transitions, and $\phi_Z$ must likewise intertwine with $g$ to preserve outputs. In a category-theoretic sense, we are demanding *compatibility of compositions* in a commutative diagram, rather than using the term "homomorphism" in a strictly group-theoretic sense.

**Definition 4.2 (observer homomorphism):** Let

$$O_1 = (X_1, Y_1, Z_1, f_1, g_1, \mathcal{B}_1) \quad \text{and} \quad O_2 = (X_2, Y_2, Z_2, f_2, g_2, \mathcal{B}_2)$$

be two observers (as in Definition 4.1). A *homomorphism* from $O_1$ to $O_2$ is a triple of functions $(\phi_X, \phi_Y, \phi_Z)$ such that:

$$\begin{aligned} \phi_X &: X_1 \to X_2, \\ \phi_Y &: Y_1 \to Y_2, \\ \phi_Z &: Z_1 \to Z_2, \end{aligned}$$

and for all $x \in X_1$ and $y \in Y_1$, the following commutation conditions hold:

$$\phi_X\big(f_1(x, y)\big) = f_2\big(\phi_X(x), \phi_Y(y)\big),$$

$$\phi_Z\big(g_1(x)\big) = g_2\big(\phi_X(x)\big).$$

Intuitively, $(\phi_X, \phi_Y, \phi_Z)$ ensures that the transition function $f_1$ and output function $g_1$ in $O_1$ map in a structure-preserving way to $f_2$ and $g_2$ in $O_2$. Equivalently, we have two commutative diagrams (see Figs. 4.10 and 4.11):

**Theorem 2 (equivalence relation):** *Let $O_1$ and $O_2$ be two observers. Define $O_1 \sim O_2$ if and only if there exists a* bijective *homomorphism $(\phi_X, \phi_Y, \phi_Z)$ between $O_1$ and $O_2$. Then $\sim$ is an equivalence relation on observers.*

**Reflexivity:** Take $\phi_X = \mathrm{id}_{X_1}$, $\phi_Y = \mathrm{id}_{Y_1}$, $\phi_Z = \mathrm{id}_{Z_1}$. These trivially satisfy the commutation conditions in both diagrams.

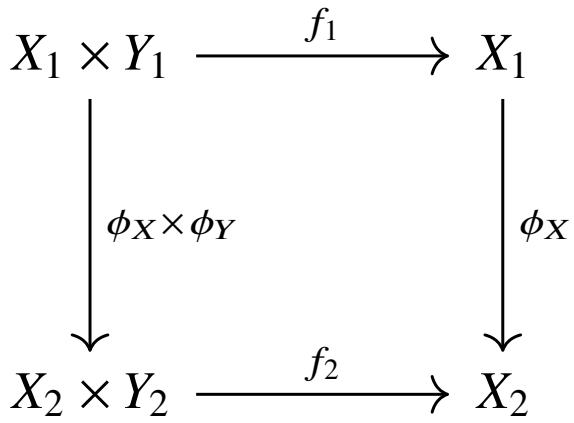

*Figure 4.10* Commutative diagram for the transition function $f$.

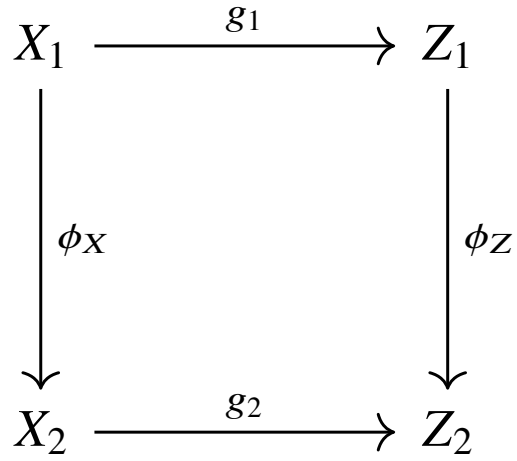

*Figure 4.11* Commutative diagram for the output function $g$.

**Symmetry:** If $(\phi_X, \phi_Y, \phi_Z)$ is a bijection from $O_1$ to $O_2$, then $(\phi_X^{-1}, \phi_Y^{-1}, \phi_Z^{-1})$ is a bijection from $O_2$ to $O_1$, satisfying the same commutative properties in reverse.
**Transitivity:** If $(\phi_X, \phi_Y, \phi_Z)$ is a bijection from $O_1$ to $O_2$ and $(\psi_X, \psi_Y, \psi_Z)$ is a bijection from $O_2$ to $O_3$, then composing them yields a bijection from $O_1$ to $O_3$ preserving all commutation requirements.

Thus, $\sim$ is an equivalence relation.

Two observers $O_1$ and $O_2$ are said to be *equivalent* if they differ only by a bijective, diagram-commuting relabeling. In category-theoretic language, this corresponds to an *isomorphism* between observer objects. Any property of an observer that remains invariant under such isomorphisms (e.g., minimal cardinalities of $X, Y, Z$, the presence of certain feedback cycles, or stable attractors) qualifies as an *invariant* of the equivalence class.

### *4.10.4.2 Complexity metrics and results*

#### OBSERVATIONAL COMPLEXITY MEASURE

To quantify the "size" or "sophistication" of an observer $O = (X, Y, Z, f, g, \mathcal{B})$, define

$$C(O) \;=\; \log\big(|X| \times |Y| \times |Z|\big) \;-\; \Lambda(O),$$

where $\log\big(|X| \times |Y| \times |Z|\big)$ captures the combinatorial capacity of internal states, inputs, and outputs, and $\Lambda(O)$ accounts for redundancies or symmetries in $f$ and $g$. (If multiple $x \in X$ respond identically and produce identical outputs, they do not increase genuine complexity.)

**Proposition 4.1 (bounds on observational complexity):** For a minimal observer with $|X| > 1,\ |Y| \geq 1,\ |Z| \geq 1$, we have

$$C(O) \;\geq\; \log(2).$$

Furthermore, if $O$ can adapt or learn (increasing $|X|$ or altering $f, g$), then $C(O)$ can grow arbitrarily large.

*Proof.* A minimal observer requires $|X|\times|Y|\times|Z| \geq 2$, so $\log\big(|X|\times|Y|\times|Z|\big) \geq \log(2)$. While $\Lambda(O)$ subtracts redundancies, it cannot push $C(O)$ below zero for a structurally minimal system. As $|X|, |Y|, |Z| \to \infty$, or as $f, g$ become more varied, $C(O)$ can increase without bound.

**LOOP EFFICIENCY AND ADAPTATION SPEED**

Lastly, consider how quickly an observer $O$ "adapts" to a given environment. Define an *adaptation function*

$$\alpha_O : X \times Y^* \;\to\; \mathbb{N},$$

which, for a sequence of inputs $(y_1, y_2, \dots) \in Y^*$, returns the time or number of state transitions required for $O$ to reach a stable configuration or fulfill a measurement/control goal. In finite-state systems, one often proves boundedness of $\alpha_O$ using Markov chain hitting-times; in continuous domains, Lyapunov methods or approximate dynamic programming may be invoked.

Optimizing $\alpha_O$ across all allowable designs of $(f, g)$ is akin to an *optimal control* or *reinforcement-learning* problem, seeking minimal expected adaptation time. Notably, two observers belonging to the same equivalence class via Theorem 2 (i.e. isomorphic) must exhibit the same adaptation profile, complexity measure, and other invariants, up to a relabeling of states, inputs, and outputs. This highlights the relational consistency emphasized previously and ensures that observer structure, rather than mere notation, dictates the system's dynamics.

### *4.10.5 Addressing counterarguments*

#### *4.10.5.1 Reductionism critique*

Some worry that labeling simple devices (thermostats) as "observers" trivializes the concept. We rebut that minimal observers are foundational building blocks: layering, second-order loops, or enriched predictive mechanisms $h$ can yield the complexity of conscious or social systems. The presence of feedback and boundary definition $\mathcal{B}$ is the *sine qua non* of observation, whether in a simple or advanced entity.

#### *4.10.5.2 Infinite regress in self-reference*

Second-order or multi-layer observers can appear to regress infinitely: who observes the observer's observer, etc.? We propose hierarchical encapsulation: each observer only references or modifies a finite subset of its own states. Formally, we forbid cycles of observation that do not converge or yield stable references. This ensures a well-founded partial order in the lattice of meta-observation relations.

## 4.11 Conclusions and outlook

We have presented a comprehensive, rigorously formalized theory of minimal observers that unifies concepts from cybernetics, quantum measurement, digital physics, and philosophical discussions of realism, meaning, and consciousness. Our model establishes the *minimal* criteria (sensing, action, state transitions, boundary) that constitute an observer, demonstrates how fundamental notions (measurement outcomes, reference frames, hierarchical organization) hinge on the presence of such observers, engages with Kantian, Husserlian, and Wittgensteinian perspectives, situating our feedback-based approach within classical philosophical discourse, offers rigorous mathematical results on observer equivalences, complexity metrics, and loop adaptation speeds, and addresses key critiques (reductionism, infinite regress) by highlighting scalability, hierarchical encapsulation, and boundary reconfiguration as essential elements. Through explicit *diagrams* illustrating core feedback loops and boundary definitions, we have shown how the observer concept can be visualized in contexts ranging from simple thermostats to multi-layer experimental apparatus in physics. In bridging computational, physical, and philosophical dimensions, this framework aspires to be a definitive, self-contained theory of observation.

In sum, recognizing *observation* as a fundamental, feedback-driven process—constrained by boundary definitions, state transitions, and sensor-actuator loops—offers a powerful lens for explaining measurement, emergent complexity, and the construction of meaning. We hope this work will inspire further efforts across disciplines to refine and adopt the minimal observer framework, illuminating the deep interweave of cognition, physics, computation, and philosophy in shaping our understanding of reality.

## Acknowledgments

We thank Stephen Wolfram for engaging discussions on observers. We also like to thank John DeBrota for useful suggestions and comments on QBism. This work was supported by the following grants: The Foundational Questions Institute and Fetzer-Franklin Fund, a donor advised fund of Silicon Valley Community Foundation [FQXi-RFP-1817]; the John Templeton Foundation [Grant ID 62106]; and the Australian Research Council [Grant DP210100919].

## Notes

1. Notably, from the perspective of Giulio Tononi's *Integrated Information Theory* (IIT), even such a minimal feedback system could be considered to have an extremely rudimentary form of consciousness, as it integrates information (the sensed temperature) and produces a differentiated response (heating action)—albeit at a very low level of complexity or $\Phi$, IIT's measure of consciousness [20].
2. A natural extension of this minimal observer into the quantum domain could leverage epistemic interpretations of quantum mechanics, such as QBism. In QBism, probabilities associated with quantum states represent an observer's subjective degrees of belief or betting odds about measurement outcomes. Thus, a quantum minimal observer might be formalized as an entity whose internal states correspond to evolving belief states updated via quantum Bayesian inference rules, driven by the outcomes of quantum measurements relative to their actions on the world.